# Design and Commissioning of an LWA Swarm Station: The Long Wavelength Array – North Arm

C. A. Taylor [1,*], J. Dowell [1], G. B. Taylor [1], K. S. Obenberger [2], S. I. Chastain [1], J. Verastegui [1], L. E. Cordonnier [1], P. Kumar [3], E. Sheldahl [1], S. Bruzewski [4], T. Dolch [5,6] and C. A. Siders [1]

[1]Department of Physics and Astronomy, University of New Mexico, Albuquerque, NM, USA

[2]Air Force Research Laboratory, Space Vehicles Directorate, Kirtland AFB, NM, USA

[3]Curtin Institute of Radio Astronomy, Curtin University, Perth, WA, Australia

[4]United States Naval Observatory, Washington, DC, USA

[5]Department of Physics, Hillsdale College, Hillsdale, MI, USA

[6]Eureka Scientific, Oakland, CA, USA

*ctaylor98@unm.edu; craig.t.wsu@gmail.com

Modern radio interferometers are designed with increasingly sprawling geographical footprints, offering enhanced sensitivity and resolution. However, managing such extensive facilities presents operational challenges that can potentially impede or delay scientific progress. One solution to such obstacles is the ‘swarm telescope’ concept which enables collaborative use of individual telescope systems, overseen by separate institutions, to create a more powerful and manageable facility. We present the design, construction, and commissioning of the Long Wavelength Array – North Arm (LWA-NA) station, a prototype 64-element LWA Swarm telescope. LWA-NA is a cost-efficient, rapidly deployable platform for radio astronomy, and serves as a pathfinder for the larger LWA Swarm project.

## 1. Introduction

Low-frequency radio astronomy has experienced significant growth with the continued development of the Low-Frequency Array collaboration (LOFAR; van Haarlem *et al.*, 2013; International LOFAR Telescope, 2023) including the completion of the New Extension in Nancay Upgrading LOFAR site (NenuFar; Zarka *et al.*, 2020), Murchison Widefield Array (Tingay *et al.*, 2013; Wayth *et al.*, 2018), the Long Wavelength Array (LWA; Taylor *et al.*, 2012; Cranmer *et al.*, 2017; Dowell & Taylor, 2018), and groundbreaking at the Square Kilometer Array Low-band array (Dewdney *et al.*, 2009; Labate *et al.*, 2017). Recent advances in computing technology have enabled contemporary interferometer telescopes to achieve greater frequency coverage, larger collecting areas, and fields of view spanning the entire sky. Efficient operation of aperture array telescopes is essential for impactful science production in an ecosystem that emphasizes maximizing the number of elements for ever greater sensitivity and resolution. In this effort, the LWA is a well-understood platform for advancing low-frequency radio astronomy, offering unique capabilities as a mid-latitude telescope, including real-time all-sky imaging at wide bandwidths, commensal observing modes, a modular design that facilitates future expansion, and a flexibility to dynamic upgrades.

*Corresponding author.

The LWA is an interferometer array telescope collaboration comprised of 'stations' distributed across the Southwestern United States. Each station is a digital beamforming aperture array capable of targeted observing and all-sky imaging on its own, or in cooperation with other LWA stations to form an interferometer. These stations observe over a frequency range of 3–88 MHz and are composed of a pseudo-random distribution of dual-polarization dipole antennas spread across a ∼100 m aperture (Taylor *et al.*, 2012). Currently, there are three operational LWA stations: LWA1 co-located with the Very Large Array, LWA-SV located at the Sevilleta National Wildlife Refuge, and one hosted at the Owens Valley Radio Observatory named OVRO-LWA (Anderson *et al.*, 2019). The first two stations, LWA1 and LWA-SV, are what the LWA Collaboration considers a 'standard' LWA station equipped with 256 elements. OVRO-LWA has had two stages of upgrades since its deployment, currently operating with a core of 243 stands and an additional 109 stands with long baselines up to ∼2.4 km for arc-minute scale imaging.

Single station observing campaigns with these three instruments have contributed to a wide variety of fields such as pulsars (Kumar *et al.*, 2022), Jovian decametric radio emission (Clarke *et al.*, 2014), meteor radio afterglows (Obenberger *et al.*, 2014; Varghese *et al.*, 2024), cosmic dawn (DiLullo *et al.*, 2020, 2021), lightning evolution (Stock *et al.*, 2023), and ionospheric phenomena (Obenberger *et al.*, 2020, Obenberger *et al.* in prep.). However, limited collaborative efforts have been made between these three operating stations due to various stages of in-progress upgrades and on-site construction. To effectively make use of the LWA as a long baseline interferometer, additional stations must be constructed to improve coverage in the $(u, v)$-plane and access arc-second scale imaging resolution.

The LWA Swarm project proposes to create a long-baseline aperture synthesis telescope by deploying configurable LWA 'swarm stations' to host universities and colleges for independent and cooperative science as part of the LWA Collaboration (Taylor *et al.*, 2020), to form an open skies low-frequency "long baseline aperture synthesis imaging array" at mid-latitudes (Taylor *et al.*, 2012). Central to this program is an already established network to interconnect individual stations with an autonomous monitor and control, observing, and data transport system through the LWA Software Library (`lsl`; Dowell *et al.*, 2012). By targeting accessibility and scalability, this project offers an alternative direction in an astronomy landscape where new instrument proposals typically come with costs, complexity, and geographical scales that are ill-suited for single institutions to manage. With the completion of the Long Wavelength Array — North Arm (LWA-NA) station, we aim to show that a six-station LWA interferometer has the sensitivity and resolution to investigate active topics in radio astronomy such as the propagation dynamics in X-ray binary jets (Tetarenko *et al.*, 2019), imaging of the emerging population of long period pulsars (Hurley-Walker *et al.*, 2023, Rea *et al.*, 2024), and survey observations of Galactic Center targets that are inaccessible to instruments such as LOFAR (de Gasperin *et al.*, 2021).

The prototype station for the LWA Swarm telescope is the Long Wavelength Array – North Arm (LWA-NA), a 64 element dipole instrument located less than a kilometer west of the end of the north arm of the Karl G. Jansky Very Large Array (VLA). LWA-NA represents what we consider an LWA 'swarm' station, a rapidly deployable and lightweight platform for radio astronomy based on the Swarm Telescope Concept described in Dowell & Taylor (2018). This concept places an emphasis on utilizing open-access and commercially available hardware to provide a cheaper entry point to radio astronomy at the university level. In this paper, we describe the design, construction, and commissioning of LWA-NA, an LWA swarm station. Section 2 provides an outline of the design considerations taken for LWA-NA, a description of the station construction, and a sensitivity estimate for LWA-NA. Section 3 illustrates the end-to-end signal chain of the system, details clock distribution, and outlines station observing modes. Section 4 describes commissioning the swarm station, from instrument control to preliminary test observations. Lastly, in Sec. 5, we discuss the implications of this new low-frequency long-baseline interferometer, additional swarm stations already under construction, and currently underway characterization observations. For the convenience of the reader, a glossary of acronyms is provided in Table 1.

## 2. Design and Construction

### 2.1. *Design*

Standard LWA station array configurations are made by pseudo-randomly distributing stands uniformly over a 100 m aperture, at a minimum separation of five meters, with a slightly elliptical shape to improve the main lobe symmetry of the synthesized beam at lower elevation angle pointings (Ellingson *et al.*, 2013). An LWA Swarm station layout applies a Gaussian taper to the array stand density to allow for a budget-adjustable station size, with the minimum recommended number of elements to provide acceptable baseline sensitivity for interferometry applications and reasonable all-sky images being 48 dipoles (Dowell & Taylor, 2018). For the LWA-NA telescope, the design specifically samples 64-antenna locations after applying this density tapering to the standard LWA station layout, to give an aperture $73.80 \times 66.75$ m in size, elongated in the North-South axis. This miniature station design enables future experimental LWA upgrades to be tested on a lower overhead system compared to the standard station size.

Table 1. Glossary of terms and acronyms.

| Term | Description |
|---|---|
| ARX | Analog Receiver subsystem |
| ASP | Analog Signal Processor |
| BFU | Beamformer Units |
| COR | Wideband Correlator for LWA |
| DR | Data Recorder |
| DRX | Digital Receiver |
| EMI | Electromagnetic Interference |
| Es | Sporadic-E |
| FEE | Front-End Electronics |
| FPGA | Field Programmable Gate Array |
| GPS | Global Positioning System |
| LWA | Long Wavelength Array; Interferometer array telescope collaboration |
| LWA1 | First station of the Long Wavelength Array located at the VLA |
| LWA-NA | Long Wavelength Array – North Arm |
| LWA-SV | Long Wavelength Array – Sevilleta; located at Sevilleta Nat'l Wildlife Refuge |
| MCS | Monitor and Control Software |
| MMCX | Micro-Miniature Coaxial Connectors |
| NDP | Next generation Digital Processor for LWA |
| Orville | Orville Wideband Imager |
| OVRO-LWA | The Owens Valley Radio Observatory Long Wavelength Array station |
| PDU | Power Distribution Unit |
| PPS | Pulse-per-second signal |
| QMA | Quick Mating connector |
| RFI | Radio Frequency Interference |
| RPD | RF and Power Distribution |
| SEFD | System Equivalent Flux Density |
| SEP | Station Entry Panel |
| Stand | A pair of orthogonally-aligned active dipoles sharing a mast |
| Station | An antenna array and associated electronics |
| TBF | Transient Bandwidth Frequency-domain |
| TPC | Timing and Power Control unit |
| UNM | University of New Mexico |
| UPS | Uninterruptible Power Supply |
| VLA | Karl G. Jansky Very Large Array |

The LWA-NA site (Fig. 1) is a $120 \times 120$ m enclosed lot located ∼0.5 km southwest of pad VLA: N72 at the terminus of the VLA north arm. Previously this location was an LWA testing site housing equipment from precursor experiments and a Low-Frequency All-Sky Monitor site (LoFASM; Dartez, 2021), thus, already equipped with a 1 Gbps fiber link to the nearby VLA pad, and an existing power connection. Throughout the development and construction, we prioritized readily available hardware to reduce overhead costs and to demonstrate a cost-effective entry point for university-scale radio astronomy. To further reduce costs, LWA equipment occupying the site prior to breaking ground for this project was deconstructed. The undamaged antenna stands and PVC conduit-shielded cable were collected for re-use in the construction of LWA-NA. Coaxial cable returned from this site was inspected for damage and tested for performance in the LWA lab at the University of New Mexico (UNM) before being reassigned to appropriately positioned stands in the new LWA-NA station layout. Recycled hardware recovered in preliminary deconstruction of the site, including cable and antenna components, comprised over 20% of the material used to build LWA-NA.

## 2.2. *Construction*

After parts acquisition, the site was surveyed using a Northwest Instruments NTS02B Total Station and reflector to designate all on-site hardware placements. Following the initial site survey, antenna masts were installed using a jackhammer and a portable generator. With the surveyed stand placements and mast locations, an approximate trenching map was created to minimize the total cable length requirements using `scikit-image` (Van Der Walt *et al.*, 2014) and can be seen in Fig. 2 (see also Dowell & Taylor, 2018, and supplementary files). For the electronics shelter, a 90 ×54 in. concrete foundation pad was poured on-site with embedded conduit to allow cable to enter the shelter underground, and a modified American Products enclosure was craned onto the foundation with the help of NRAO/VLA staff. A walk-behind trencher was used to run fiber and power to the shelter, and connections to our suite of station monitoring electronics comprised of a weather station (Davis VantagePro2), GPS (Spectracom SecureSync 1200–233), and lightning detector (Boltek EFM-100C).

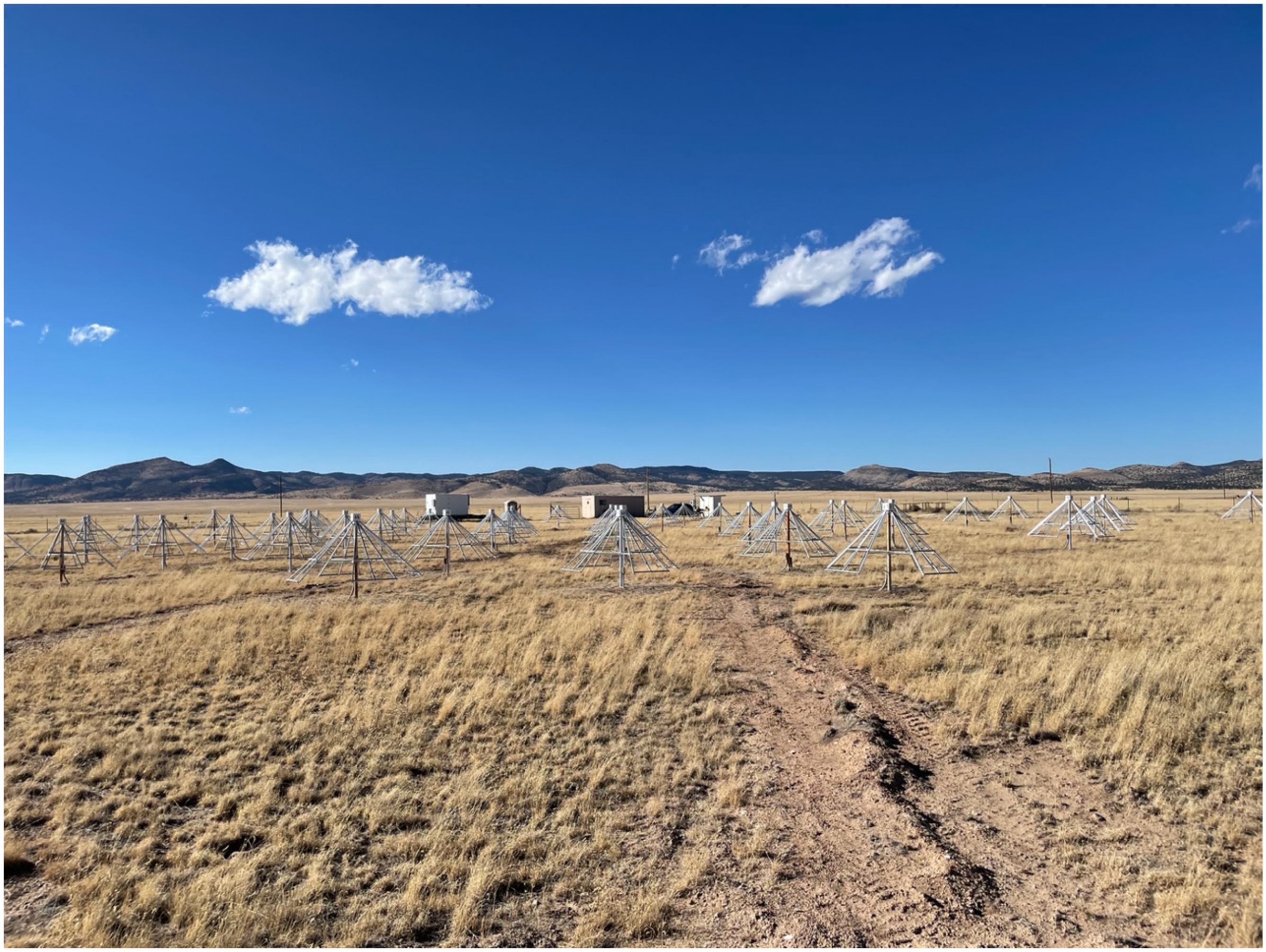

Fig. 1. (Color online) Completed LWA – North Arm station viewed from the Southeast corner of the array. Objects in the background from left to right: Mobile Trailer for covered work, storage container, work vehicle, LWA-NA Electronics Shelter.

Array trenches were cut to a depth of 18–24 in. to preserve the thermal stability of coaxial cable (LMR-195) used underground from the shelter into the array (Craig, 2009b). Upon completion of the main trenching, cable was laid to each antenna, with excess cable wound into a large auxiliary trench adjacent to the electronics shelter, before backfilling the trenches and burying the cable. Next, the antenna assemblies were constructed by a combination of graduate and undergraduate student workers, and the electronics racks were populated with back-end hardware necessary for the station. An 'As-Built' site survey was performed after the installation of all hardware to accurately determine the positions of each stand using the Northwest Instruments total station, and final cable lengths to each antenna were measured using time-domain reflectometry. In late 2023, the swarm station was officially on the sky with first light; a screenshot of the new Channel #4 of the LWA-TV[a] in Fig. 3 shows the typical sky seen by LWA-NA. The construction from bare ground to the assembled station took approximately 18 months to complete (May 2022–November 2023), although there were numerous delays due to lingering post-pandemic supply chain issues and a lack of convenient housing near the LWA-NA site. We estimate that swarm stations could be deployed in as little as a few months with all parts in hand before beginning construction.

The following subsection describes the LWA-NA electronics shelter and modifications required for use in radio astronomy applications.

[a] https://leo.phys.unm.edu/lwa/lwatv4.html

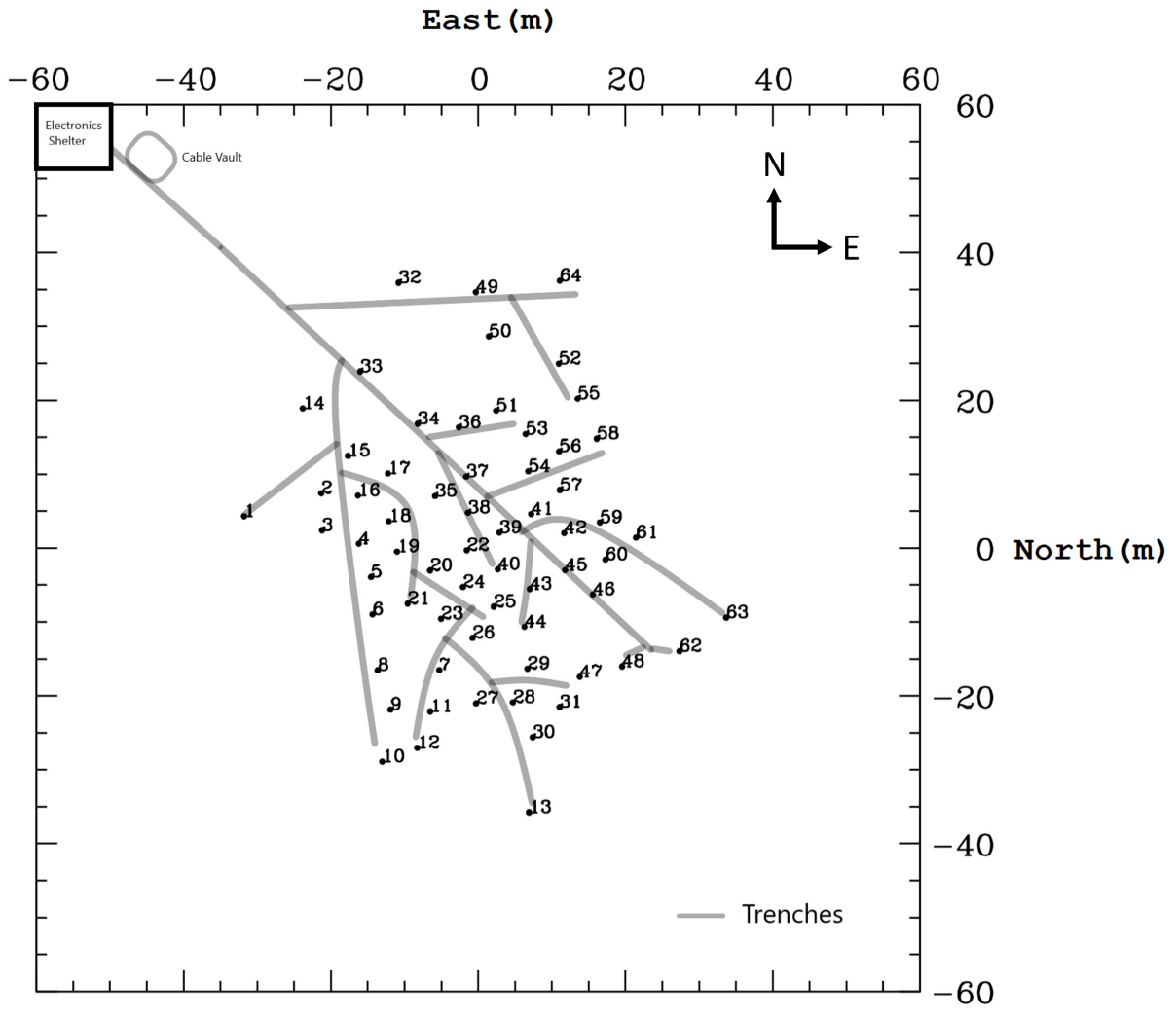

Fig. 2. Diagram of the LWA North Arm site as constructed. Antenna positions were distributed using an optimization code (see LWA Memo #210 and supplemental materials), gray lines indicate where trenches were dug during the construction phase, and the plot axis borders effectively serve as a reference to the fence-line enclosing the site. Note that a temporary outrigger fixture is located outside the fence approximately 250 m due Southeast of Stand #58 for future calibration testing.

## 2.3. *Electronics shelter*

LWA1 and LWA-SV house their back-end electronics in a modified $8 \times 9.5 \times 20$ ft. cargo container, with approximate shielding after modifications of $\sim 100$dB from antenna to electronics. The primary changes to improve shielding in these containers included installing EMI-shielded doors (60 dB attenuation rating), shielded electronics racks, HVAC ducting with honeycomb filters, continuous welded seams, and a built-in shelter entry bulkhead (Craig, 2009a, Dowell, 2022). This style of electronics shelter is expensive and excessively large for a quarter-sized LWA station considering the LWA Swarm's aim to provide a cost-effective LWA station model. Instead, the LWA-NA uses a commercially available $78 \times 90 \times 42$ in. (126 RU) American Products (AmPro) Freedom Series 3-bay telecommunications enclosure. Two of the three bays are connected and serve to house all of the back-end electronics (we refer to this joined compartment as the 'two-bay'), while the third bay is isolated from the others by a pair of center panels to house the station entry panel (SEP) for array inputs. The northward-facing doors of the connected main electronics bays each have a door-mounted 20,000 BTU cooling capacity HVAC system (total of 40,000 BTU for the two-bay).

This model of AmPro shelter is designed to be weather resistant but includes little mitigation for RF leakage throughout the enclosure. To improve the electromagnetic shielding of the shelter, several modifications were made before transporting the unit to the site. Four of the six enclosure doors and each interior auxiliary panel required removing paint and the manufacturer's water-resistant gasket to replace with a conductive water-resistant gasket, steel mesh honeycomb screens were installed over the HVAC air supply and return registers, and additional holes between bays were filled using steel wool and copper tape. TE Connectivity 30VK6C electromagnetic interference (EMI) filters were installed for the 120 V outlet power inputs, the 240 V HVAC power, and the rack electronics power lines to protect against unintended EMI leaking out along power lines. The UNM Machine Shop modified the two manufacturer center panels, dividing the back-end electronics two-bay from the SEP bay, to serve as bulkheads for antenna inputs and to facilitate the site weather station, lightning detector, GPS, and on-site internet connections (conductive gasket was also installed on these panels).

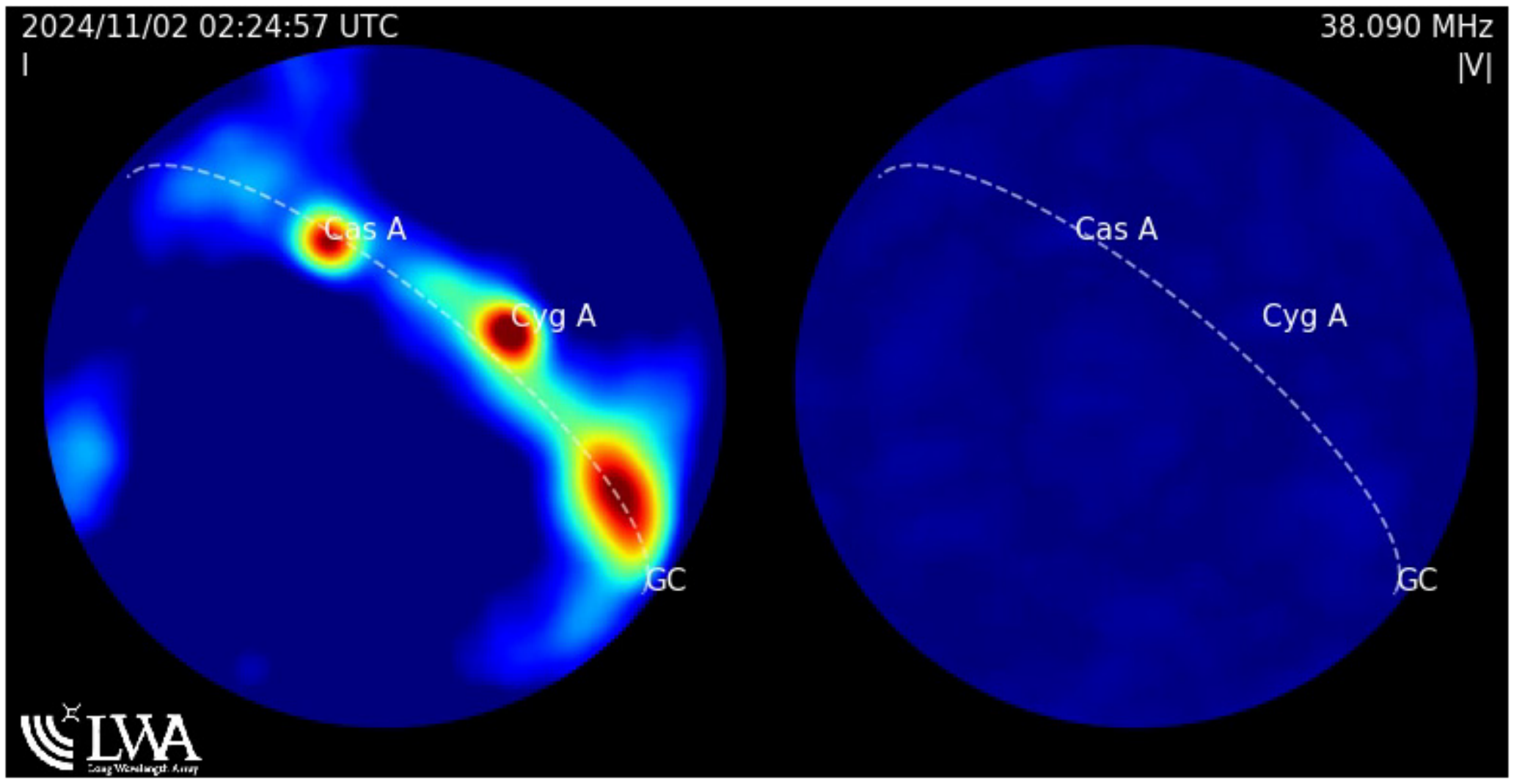

Fig. 3. (Color online) A 'typical' sky image from the LWA – North Arm prototype swarm station with Stokes-I (left) and Stokes-V (right) all-sky images. Image taken from the LWA-TV4 on November 2nd, 2024.

The shielding level of this modified enclosure was measured using a Diamond D130J discone antenna and FSH3 spectrum analyzer placed within the two-bay of the shelter, transmitting a broadband signal. Average RF power was measured using a Create CLP-5130-2N directional log-periodic antenna, located 3 m from the enclosure, and calibrated to measure the average power received as 0 dB at 120 MHz with the enclosure doors opened. With the doors closed, the average power received measured approximately −40 dB at 120 MHz, while the shelters used at LWA1 and LWA-SV are specified for −100 dB. Shielding measurements were conducted outside the Physics and Astronomy building at UNM where the enclosure was delivered and modified, and at the LWA-NA site after deployment, with both showing comparable results. This level of shielding was deemed acceptable given that extensive and costly alterations would be required to markedly improve the RF insulation of this commercial enclosure (Craig, 2009a; Taylor *et al.*, 2023). Measurements of the shielding below 100 MHz would have been preferred, however, it proved difficult to acquire an antenna that could broadcast closer to the LWA observing window and fit within the enclosure under test. Likewise, acquiring a larger and more directionally sensitive log-periodic antenna could have improved these measurements, but these antenna are not practical for handheld measurements and would require additional accommodations outside the scope of this work.

## 3. Deployment

In this section, we will describe the signal chain of the LWA-NA station from incident radiation to science outputs for both the analog and digital systems. Next follows an outline of timing distribution using the LWA-NA Timing and Power Control module, and a brief description of LWA station calibration techniques. Lastly, we will discuss the various observing modes at LWA-NA and their context compared to other LWA stations. The signal path for LWA stations has been documented in several publications and LWA memos, thus, we will briefly summarize the end-to-end signal chain for clarity and provide Fig. 4 as a visual reference for the LWA-NA architecture specifically.

### 3.1. *Analog*

The analog signal path consists of four parts: front end electronics (FEE), RF and power distribution (RPD), shelter entry panel (SEP), and analog receiver (ARX) subsystems.

Incident radiation is detected at the FEE board housed at the apex of LWA stands by measuring the differential induced voltage between two antenna arms. FEEs then assimilate and amplify the signals by 36 dB with a small portion of that gain amplification to offset the cable loss in transmission (for more details on LWA FEEs see Ellingson *et al.*, 2013). The RPD subsystem transports signals along low-loss coaxial cable (LMR-195) underground, and through buried conduit embedded in the concrete pad into the station cable entry bay. RPD also supplies FEE power via bias-Ts in the ARX subsystem and is carried on the same buried coaxial cables to each antenna. With two polarization feeds per antenna, the 128 N-type inputs pass through a lightning surge arrestor stage (Polyphaser GT-NFM-AL), then through the custom bulkhead SEP panel into the two-bay. Inside the two-bay, signals are carried on RG316 coaxial cable and utilize micro-miniature coaxial connectors (MMCX) or quick mating (QMA) connectors for efficiency in installation and testing.

Interior to the SEP bulkhead, the incoming signals are carried into the shielded input doors of the Analog Signal Processor (ASP) module, then input to one of eight Revision H ARX boards (D'Addario, 2019, 2020b). The Rev. H ARX boards provide adjustable gain of up to 63 dB in available steps of 0.5 dB, but nominally ASP control uses 2 dB steps to conform with the original specification for LWA1. Initial filtering is employed by a pair of analog filters composed of two high-pass filter and two low-pass filter options to appropriately address undesired signals at the band edges (all frequencies indicate the filter 3 dB point unless otherwise stated). The high-pass filter options are 18.7 MHz to attenuate daily strong reflected emission from the ionosphere or a through connection effectively giving −3 dB at 3 MHz for ionosphere-specific observing applications, while the low-pass filters are at 82.5 MHz to attenuate the FM radio band or 75.8 MHz which provides an earlier and smoother roll-off at the higher frequencies for Cosmic Dawn observations. Using these bandpass filter options, stations normally operate in prescribed ‘analog filter’ settings that set the ideal gain configuration for each filter pair to reduce local radio frequency interference (RFI) and provide a smooth response across the science band ($20 \sim 80$ MHz) for a broad range of applications. These configuration settings are communicated using digital signals into the subsystem for initial set-up, but remain inactive during routine operation unless specific monitor requests are made. An updated description of LWA analog filter settings for the Rev. H ARX system can be found in D’Addario (2019) and D’Addario (2020b).

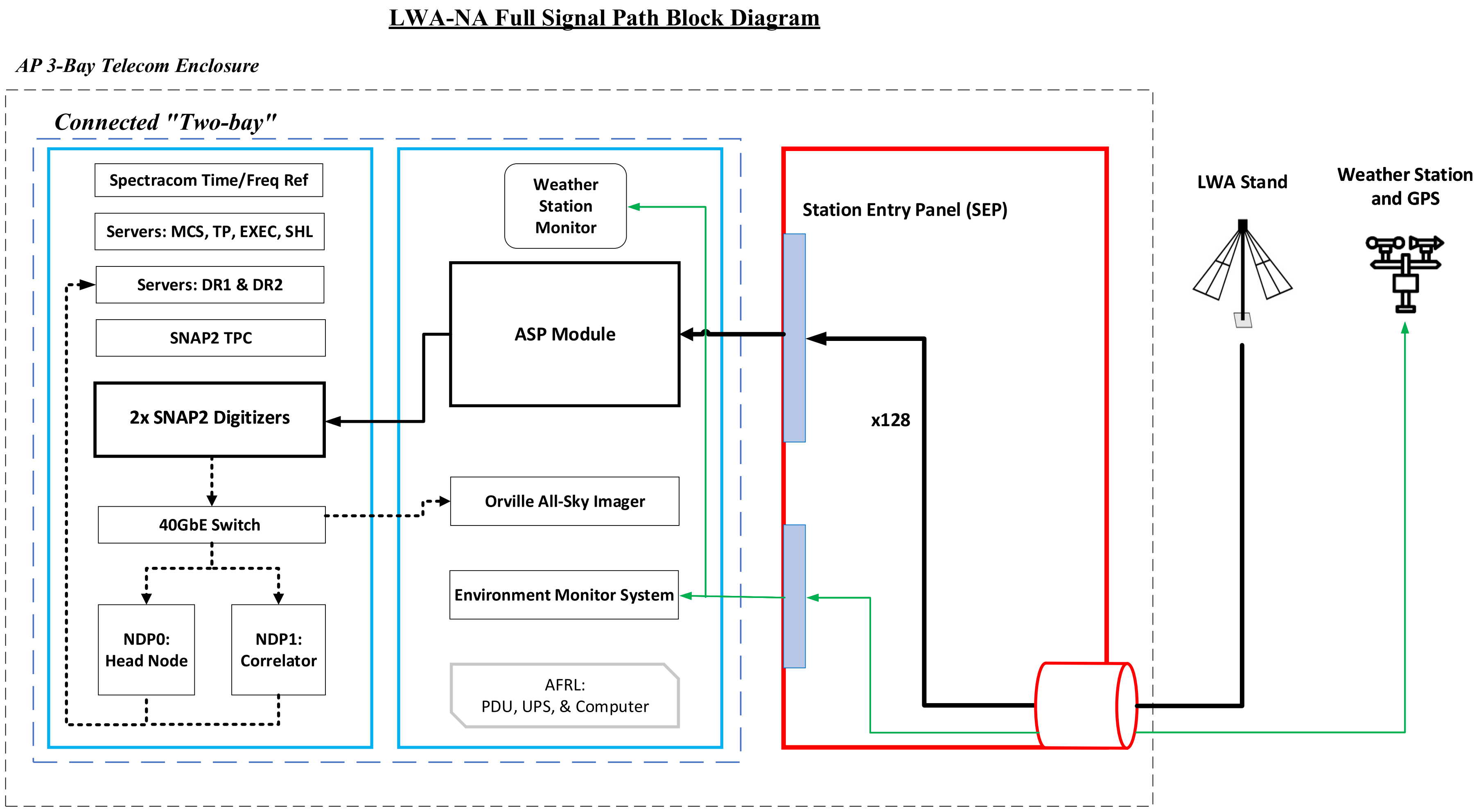

Fig. 4. (Color online) This diagram is a representation of the Electronics Shelter and enclosed subsystems for the LWA-NA end-to-end signal chain. The exterior dashed box encloses the AmPro telecommunications shelter as inputs enter through conduit. The next interior light blue box bounds the two connected main electronics bays called the "two-bay". Bold blue and red rectangles indicate the individuals bays themselves with color indicating that the two-bay is air-conditioned, while the bay containing the SEP is not cooled. Analog signals travel along solid black lines and digitized signals follow the dotted paths, from measurement at LWA stands to unicast distribution of Fourier domain data. Major components are labeled in each bay but some connections are omitted for clarity in illustrating the signal chain. Not pictured in the two-bay are 2x APC Power Distribution Units (PDUs), 2x switches (ARUBA 1930, Arista DCS-7060CX-32S) for interal and external ethernet communication, one APC Uninterruptible Power Supply (UPS), one fire alarm, and a water intrusion detector monitored by the EMS device.

The ARX boards are housed in a custom RF-shielded case manufactured by Premier Enclosures and seated in a modified server rack. The ARX case measures $30.5 \times 24 \times 28$ in., with 14U capacity and doors on opposing sides serving as input and output bulkheads. Each removable door was fitted with 128 QMA connectors in a mirrored configuration to allow signals to follow a straight path through the ARX subsystem without crossing over the enclosure. The ARX case contains no active digital electronics by design, the only other input ports to the ARX subsystem are an RS485 port and line-filtered input AC power connections through the top of the shielded enclosure. The subsystem is monitored and controlled by a half-duplex RS485 bus and powered at 6V by a power supply mounted underneath the ARX board chassis. This internal power supply also provides the 15VDC needed to power array FEEs, via Bias-Ts on each ARX channel such that the FEEs can be toggled off independently from the ARX power. The RF-only output signals exit the back door of the ASP module and complete the analog signal path at input into the digital system on the northwest side of the two-bay.

### 3.2. *Digital*

The digital signal path comprises the Next generation Digital Processor (NDP), data recorders (DRs), and their timing and monitor control system described in the following subsection.

SNAP2 boards act as the F-engine for the NDP system at LWA-NA using Kintex Ultrascale platform field programmable gate arrays (FPGAs), first implemented in the digital system of the Stage-3 upgrades at the ORVO-LWA station (see Hickish (2023) for more details on LWA SNAP2 implementation). Each of these FPGAs performs 4096-channel FFT processing for 64 digitized inputs from the array received via two FMC terminals sampled at 196 MHz on each board. Although there is only a single provider of the SNAP2 boards, they are more cost-efficient per channel when compared to the custom digital system at LWA1 or the now defunct ROACH2 boards used at LWA-SV, and fall more in line with our emphasis on readily available hardware (Craig, 2009a; Dowell & Taylor, 2020). For 128 antenna signals, the LWA-NA station requires two SNAP2s to process the full array data streams.

Analog signals from the ASP subsystem are digitized on analog-to-digital conversion (ADC) boards, henceforth ‘digitizer’ boards, designed by collaborators at the California Institute of Technology to serve as the daughterboards to the SNAP2s (D’Addario, 2020). Each digitizer uses four Texas Instruments ADS5296A ADCs to perform conversion on 16 input signals to 10-bit precision. To fully utilize the SNAP2 interface, two digitizer boards are stacked at each SNAP2 FMC slot to facilitate 32 input data streams per terminal. The F-engine pipeline utilizes a 4096-channel polyphase filter bank with a 4-tap Hamming window on the full incoming 98 MHz Nyquist band before selecting the 73.5 MHz bandwidth output for the digital pipelines. The F-engine digital response from the SNAP2s is requantized using equalizer coefficients to better sample the 4+4-bit complex integer output, before being UDP unicast to the Digital Receiver (DRX) pipelines at a rate of 4.4 GB/s.

LWA-NA has two DRX pipelines, each receiving half of the 73.5 MHz bandwidth supplied by the SNAP2s for all 128 antennas, to perform digital processing for the main LWA science data products. Each DRX pipeline passes its ∼36.75 MHz received bandwidth between four Beamformer Units (BFUs) to conduct the digital phased array beamforming on the incoming data stream using a record of pointing coefficients relayed from the Monitor and Control Software. A phase-and-sum architecture multiplies signals using *a priori* station geometry relative to the BFU pointing per antenna, before summation to form the two orthogonal polarization products as X (E-W dipole) and Y (N-S dipole) in 32+32-bit complex floating point numbers. The output data stream is then supplied to the T-engine for further repackaging into the standard LWA DRX data structure.

The Wideband Correlator (COR) module rides along with the BFUs, but takes the raw input voltage stream and unpacks it into 8+8-bit complex integers, averages up to 95.7 kHz channel width, then computes the full baseline and autocorrelation products using a Bifrost (Cranmer *et al.*, 2017) wrapper around the Tensor Core Correlator (Romein, 2021). The 32+32-bit complex integer output of the COR block is sent via UDP at 37.8 MB/s rate to be gridded and imaged by the Orville Wideband Imager system to form the all-sky and LWA-TV movie data products. To balance the data rate of the Orville system, it first computes spectra at each antenna for station monitoring purposes, before down-selecting the full frequency resolution for all-sky imaging to ∼19.9 MHz bandwidth. The center frequency set for the LWA-NA Orville images is selected to best match the frequency range of all-sky imaging at LWA1 and LWA-SV, which generally is ≈ 38.1 MHz.[b]

The Transient Buffer Frequency-domain (TBF) mode is also commensal with the DRX pipelines receiving the output from both pipelines to recompose the full 73.5 MHz bandwidth in a 5 s ring buffer at 41.8$\mu$s resolution. Complex spectra are compiled using a subset of requested data (< 5 s) from this ring buffer for each antenna as 4+4-bit complex integers to be recorded locally on NDP.

On the NDP head node, each T-engine pipeline aggregates beamformed products from two BFUs, one from each DRX pipeline to get the full SNAP2 frequency output, for one of four independent beams to be processed into the final product supplied to the observer. The 32+32-bit frequency-domain digital outputs from DRX are zero-padded from 73.5 MHz to 98.0 MHz then a Fourier transform resampler adjusts the channel widths to 50 kHz to match the frequency structure at LWA1 and LWA-SV. Next, the 98 MHz product is down-selected to the observation defined 2 × 19.6 MHz tunings, converted back to the time domain with an inverse Fourier transform, fractional delay errors are corrected using a phase rotator, and lastly, a finite impulse response (FIR) filter is applied to help dampen aliasing at band edges. Filtered time-domain data are requantized to 4+4-bit complex integers, then the raw voltage time-series is packetized for delivery to the DRs at a maximum output of 75 MB/s (rate can be smaller for lower bandwidth modes). A flow chart summarizing the digital signal path and the LWA-NA pipeline configuration is provided in Fig. 5.

[b]LWA1, LWA-SV = 38.1 MHz; LWA-NA = 38.09 MHz.

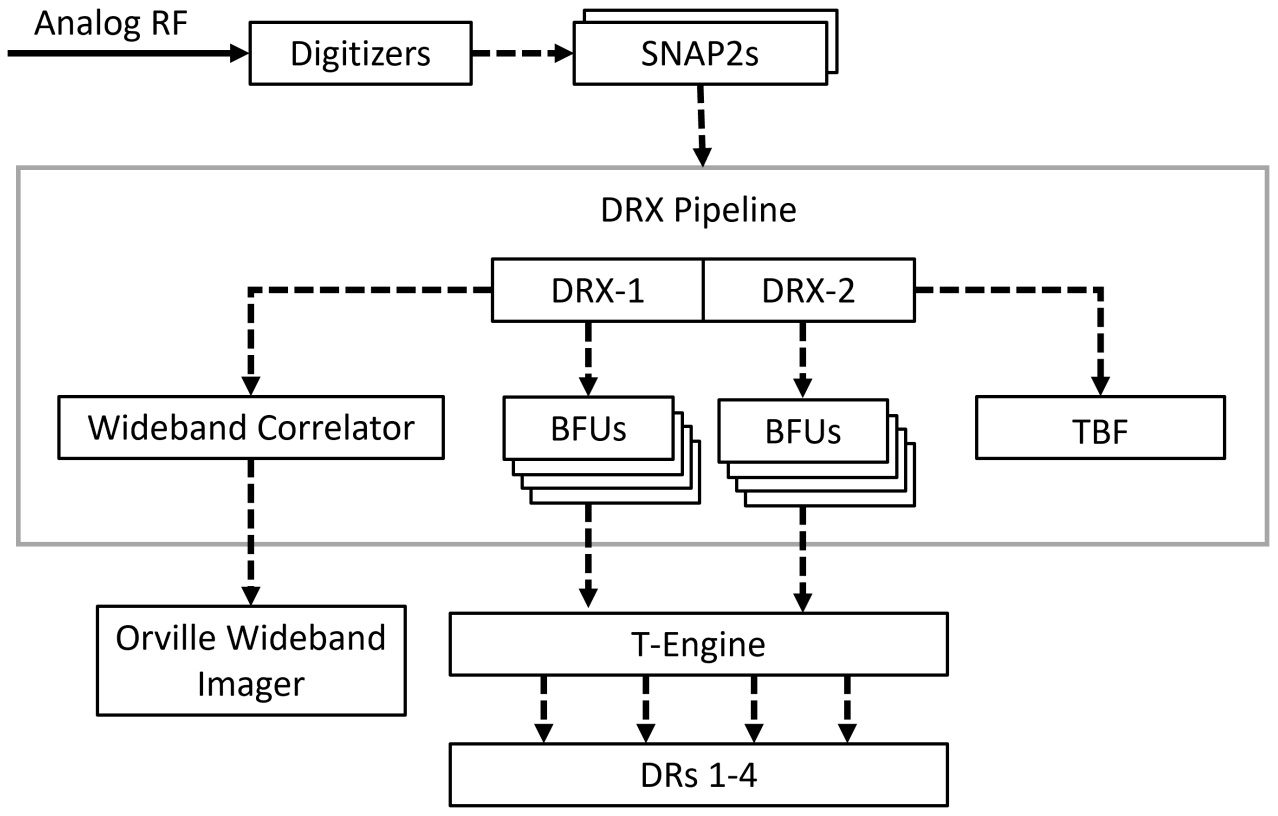

Fig. 5. Flow chart of the digital signal path for LWA-NA. The black solid line denotes the incoming analog input from the ARX subsystem and subsequent dashed lines represent digitized signals. Note that the Wideband Correlator and TBF processes receive data from both DRX pipelines (∼36.75 MHz of bandwidth for all antennas from each) indicated by the split nature of the DRX-1 and DRX-2 blocks.

### 3.3. *Timing and power control*

Timing and power distribution at the LWA-NA station for the SNAP2s and digitizers is facilitated through the Timing and Power Control module (TPC). For timing, the module receives a 1 pulse-per-second (PPS) reference signal and a 10 MHz clock signal as a phase-locked pair from The Spectracom 1200–233.[c] GPS unit mounted on the weather station adjacent to the electronics shelter. This unit is a multi-GNSS referenced rubidium oscillator model with specifications for 1 parts-per-trillion stability on the 10 MHz Frequency output and a ±25 ns accuracy when locked on the 1 PPS output. The 1 PPS is supplied to the head SNAP2 board to create a master synch pulse that is fed into the TPC module, before a fanout line driver distributes this pulse to each SNAP2. For a swarm station of this architecture, the master pulse is only fanned out to two total SNAP2s. The 10 MHz GPS output is used to synthesize a 196 MHz sample clock using a Valon 5009 dual-frequency synthesizer and passed through an 8-port splitter to the digitizers to form the required F-Engine channels. Power for the SNAP2s is distributed by a +12V internally mounted power supply through cables coming out of the front panel of the TPC module with Molex connectors.

[c]https://safran-navigation-timing.com/document/user-reference-guide-securesync-1200-time-and-frequency-synchronization-system/

### 3.4. *Station calibration and sensitivity*

Standard LWA calibration techniques correlate array dipoles with an outrigger antenna located several hundred meters from the core. These long baselines are largely insensitive to much of the diffuse emission that normally dominates the sky temperature leaving only bright point sources, namely Cygnus A and Cassiopeia A. In some cases this process also ignores baselines shorter than 183.5 m ($30\lambda$ at 49 MHz) to further aid in diminishing contributions from Galactic Plane emission (Dowell *et al.*, 2017). Discrete point sources may be isolated on these correlated long baselines by their time-variant phase due to the rotation of the sky. This single-direction calibration is applied iteratively across the LWA frequency range for individual antennas in the array through direction-independent delay adjustments. More detailed descriptions of this procedure can be found in Ellingson (2011), Ellingson *et al.* (2013), and Dowell *et al.* (2017). LWA-NA currently operates without access to an outrigger antenna, so instead we employ a diffuse calibration method using short baselines. TBF captures are correlated for baselines shorter than $4\lambda$ at a given frequency, then phase-only self-calibrated against simulated data on the same baselines using the Low-Frequency Sky Model (LFSM; Dowell *et al.*, 2017) as the reference sky model. Calibration solutions are then applied by way of adjustment coefficients to the base cable delay, per antenna, as determined from time-domain reflectometry measurements.

The performance of radio telescopes is conveniently compared using the system equivalent flux density (SEFD), given by Eq. (1), by encapsulating both aperture and instrumental sensitivity effects.

$$\mathrm{SEFD} = \frac{2k_b T_{\mathrm{sys}}}{A_{\mathrm{eff}}} \times 10^{26}\,\mathrm{Jy}. \tag{1}$$

Here, $k_b$ is the Boltzmann constant, the system temperature ($T_{\mathrm{sys}}$) is approximated by the power law relation in Eq. (2), and the effective collecting area of an LWA station ($A_{\mathrm{eff}}$) is given by Eq. (3).

$$T_{\mathrm{sys}}(\lambda) \approx 50\lambda^{2.56}\ \mathrm{K}, \tag{2}$$

$$A_{\mathrm{eff}} = N_{\mathrm{dip}} * \xi G(\lambda) \frac{\lambda^2}{4\pi} \cos^{1.6}(\theta)\mathrm{m}^2. \tag{3}$$

Here, $N_{\mathrm{dip}}$ is the number of dipoles in the array, $\xi$ is the impedance mismatch factor, $\lambda$ is the observed wavelength, $\theta$ is the zenith angle of the beam center, and $G(\lambda)$ is the antenna zenith gain ranging from 8.5 dB at 20 MHz to 5.9 dB at 88 MHz Ellingson *et al.* (2009). A recent study of the impedance mismatch factor of the LWA antenna and FEE at LWA1 and LWA-SV measures this value to be between 0–0.6 in the LWA science band (DiLullo *et al.*, 2023).

For a beam pointing sufficiently close to zenith at 74 MHz, the gain $G(\lambda)$ is approximately 6 dB, and we take the impedance mismatch factor $\xi \approx 0.5$. Thus, the idealized station SEFD is 32.3 kJy for an LWA Swarm station ($N_{\mathrm{dip}} = 64$) and 9.2 kJy for a standard LWA station ($N_{\mathrm{dip}} = 256$). Furthermore, the expected thermal noise of a given observation as determined by the SEFD of the system is provided by Eq. (4), where $n_p$ is the number of polarizations, $\Delta\nu$ is the observed bandwidth, and $t_{\mathrm{acc}}$ is the observation accumulation time. Using the above estimates of the station dependent SEFD, for a 10 min dual polarization DRX observation with an effective bandwidth of $\Delta\nu_{\mathrm{eff}} = 16.0$ MHz, the approximate thermal noise is 0.97Jy/beam for $N_{\mathrm{dip}} = 64$ and 0.24 Jy/beam for $N_{\mathrm{dip}} = 256$.

$$\Delta S = \frac{\mathrm{SEFD}}{\sqrt{n_p \Delta\nu t_{\mathrm{acc}}}}. \tag{4}$$

### 3.5. *Observing scheme*

The three main observing modes at LWA-NA are phased array beamforming utilizing the DRX pipelines, the Orville Wideband Imager, and TBF captures. These were previously introduced in the context of the digital signal chain, but here we briefly outline their data products for the end user, the contents of which are also summarized in Table 2. We also note that the LWA is an open skies observatory and the LWA-NA station will be included for user observing proposals starting in LWA Observing Cycle 14 for 2026 contingent on funding.

Table 2. Summary of LWA-NA operations specifications .

| Specification | As built description |
|---|---|
| *DRX*: | |
| Observing Mode | Phased-array beamforming |
| Beams | 4 (dual pol.) |
| Frequency Range | 3–88 MHz |
| Integration Time | >51.02 ns |
| Instantaneous Bandwidth | $2 \times 19.6$ MHz |
| Beam FWHM | $3.2^\circ$ (74 MHz) |
| *Orville*: | |
| Observing Mode | Wideband all-sky imager |
| Frequency Range | 30.0–49.9 MHz |
| Instantaneous Bandwidth | 19.9 MHz |
| Integration Time | 5 s |
| Image Shape | $120\times120$-pixel ($\sim1.083^\circ$/pixel at zenith) |
| $\mathrm{FOV}_{\mathrm{eff}}$ | $130^\circ$ |
| *TBF*: | |
| Observing Mode | Total-bandwidth antenna complex spectra |
| Frequency Range | 15–88 MHz |
| Instantaneous Bandwidth | 73.5 MHz |
| Integration Time | 41.8 $\mu$s–5 s |

DRX beamforming is the primary directed observing mode used for LWA science. LWA-NA is equipped with four independent beams, each providing two available tunings of 19.6 MHz bandwidth that can be selected at any center frequency within the station observing band. As mentioned above, baseband voltage time-series data (51.02 ns minimum cadence) is the output product of DRX beamformed observations and a suite of tools for reducing such data is available through `lsl`. Due to NDP and the SNAP2s supplying the full 98 MHz Nyquist band to the T-engine, these beams are independent within the station-configured bandwidth, albeit zero-valued outside the selected 73.5 MHz science band. Associated with the DRX pipelines, there is also the DR Spectrometer mode in which the DRs continuously perform a configurable window-sized FFT on incoming data from DRX for a time-averaged frequency domain product.

The Orville Wideband Imager operates commensally with beamed observations and, as mentioned above, is set to a center frequency of 38.09 MHz to best match the all-sky imagers at LWA1 and LWA-SV. These all-sky images captured at a 5-second cadence have a horizon-to-horizon beam ($\sim130^\circ$ effective size) and are imaged onto a coarse grid of $120 \times 120$ pixels to operate in real-time ($\sim1.083^\circ$/pixel at zenith). Orville data are stored for four weeks at 95.7 kHz channel resolution for target of opportunity searches in the all-sky images, before being averaged up to 3.25 MHz channels for archiving.

TBF captures are another available data format used for station monitoring or in special observing modes, such as for lightning self-triggered observations during nearby electrical storms (Stock *et al.*, 2023). These data are complex spectra computed for each antenna covering the 73.5 MHz band currently supplied by the SNAP2s in short durations at 41.8 $\mu$s cadence in either manual, scheduled, or triggered captures.

## 4. Commissioning

Commissioning exercises were divided into two categories to validate the construction and operation of LWA-NA: (Sec. 4.1) Station Control and Function, to verify remote interaction with LWA-NA conforms with other operational stations and that the hardware performs as expected. Section 4.2 Pointing and Observing Validation, to ensure the station can point with acceptable speed and accuracy for time-domain astronomy, and to make benchmark observations demonstrating the station is ready for use in the developing LWA Swarm.

### 4.1. *Station control and function*

RF signal path testing throughout the system was done by toggling each antenna input individually to ensure the antenna-based signal paths were correctly mapped throughout the system. This was done using the MCS software, effectively testing the hardware connection health and the mapping of data pathways from FEE to digitizer. Monitoring of station control logs and 5 s real-time digital spectra retrieved from the SNAP2 interface that post to the LWA Operations website for visual comparison are used to verify control commands were executed properly. This real-time view of digitized spectra from LWA stations is regularly used for remote or on-site signal chain troubleshooting during commissioning and for assessing LWA station health on a per-antenna basis. We also examined the spectra for gain spurs induced by the interleaved 4-channel ADCs on the digitizers, but no spur channels were found to exceed the overall average noise level of spectra.

Diurnal station temperature was measured using four 6 h DRX observations at a zenith pointing to compare with the simulated drift curve, 10 min gaps were left between each session to ensure station health. A 24 h drift curve from the LWA-NA station was then simulated in `lsl` using the LFSM with a zenith beam pointing using the post-construction static station measurement information base initialization files. We find good agreement between our observed results and the simulated data as seen in Fig. 6. Concurrent with this zenith beam, we also observed using two custom beams in the DR real-time spectrometer mode for single antennas. Stands 33 and 63 were used for this test due to their relative proximity in the array, and because they lie along a line of sight to the electronics shelter to probe for self-generated EMI. In these observations, we find similar RFI environments between all three beams and agreement with the background sky diurnal cycle.

### 4.2. *Beam pointing*

Three tests were employed to understand the performance and accuracy of the beam pointing function at LWA-NA. The first tested system communication stability and consistency by observing using all four beams simultaneously for 30 min at Cassiopeia A on April 7, 2024. We found the beams to have an effectively identical response as expected given that all beamforming is software-defined. Next, we made beam-transit observations by allowing Cassiopeia A to transit through two adjacent stationary topocentric beam pointings. Beams 1 and 3 were used in this test observation and pointings were separated by 32 min to give a slight overlap at 74 MHz between the FWHM of each beam (based on the largest antenna spacing along the East-West axis at LWA-NA of $d = 66.75$ m). Results showed accurate timing of source arrival within each beam and displayed the expected Gaussian profiles given the shape of the LWA-NA beam.

Lastly, our transient response test aimed to establish the speed at which LWA-NA can re-point the beams accurately. This was done by swapping a beam between a bright radio source, Cassiopeia A, and the North Celestial Pole which is largely a dark field at these frequencies. A starting cadence of $\Delta t = 2$ s was used for two source swaps (on-off-on-off), before geometrically reducing the point spacing by a factor of two to a final swap time of $\Delta t = 62.5$ ms. This was done for both equatorial and topocentric coordinate scheduling styles for DRX beam observations. In each case, LWA-NA successfully got onto the source within a single 62.5 ms time-step regardless of the cadence between pointings as seen in Fig. 7, with a uniform arbitrary power level at each source for all target changes. This is an encouraging result for abstract observing strategies, for example, characterizing the local ionosphere could be done with the LWA by quickly switching between many sources across the sky to build a profile of overhead scintillation. This technique can act as a form of radio frequency "speckle imaging" to create snapshots of ionospheric fluctuations for modeling purposes and data correction.

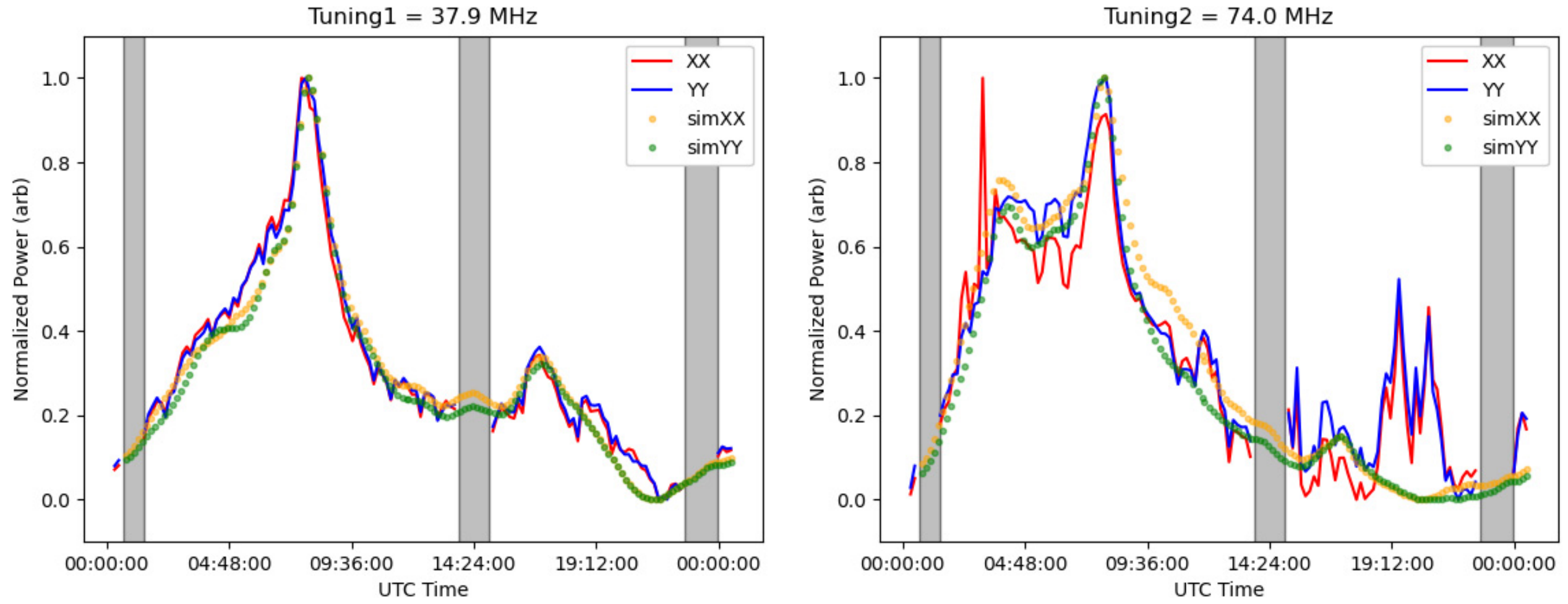

(a)

(b)

Fig. 6. (Color online) Comparison of LWA-NA diurnal brightness temperature variation (solid lines) with the simulated drift curve of the same observation using the Low Frequency Sky Model (dotted curves). Both the 37.9 MHz (a) and 74.0 MHz (b) tunings have a total bandwidth of 19.6 MHz, averaged into a single channel after flagging the data for RFI using spectral kurtosis. There is good agreement between data and the simulation throughout the observation, with exception to the enclosed gray regions where the sky is dominated by local RFI caused by power-lines.

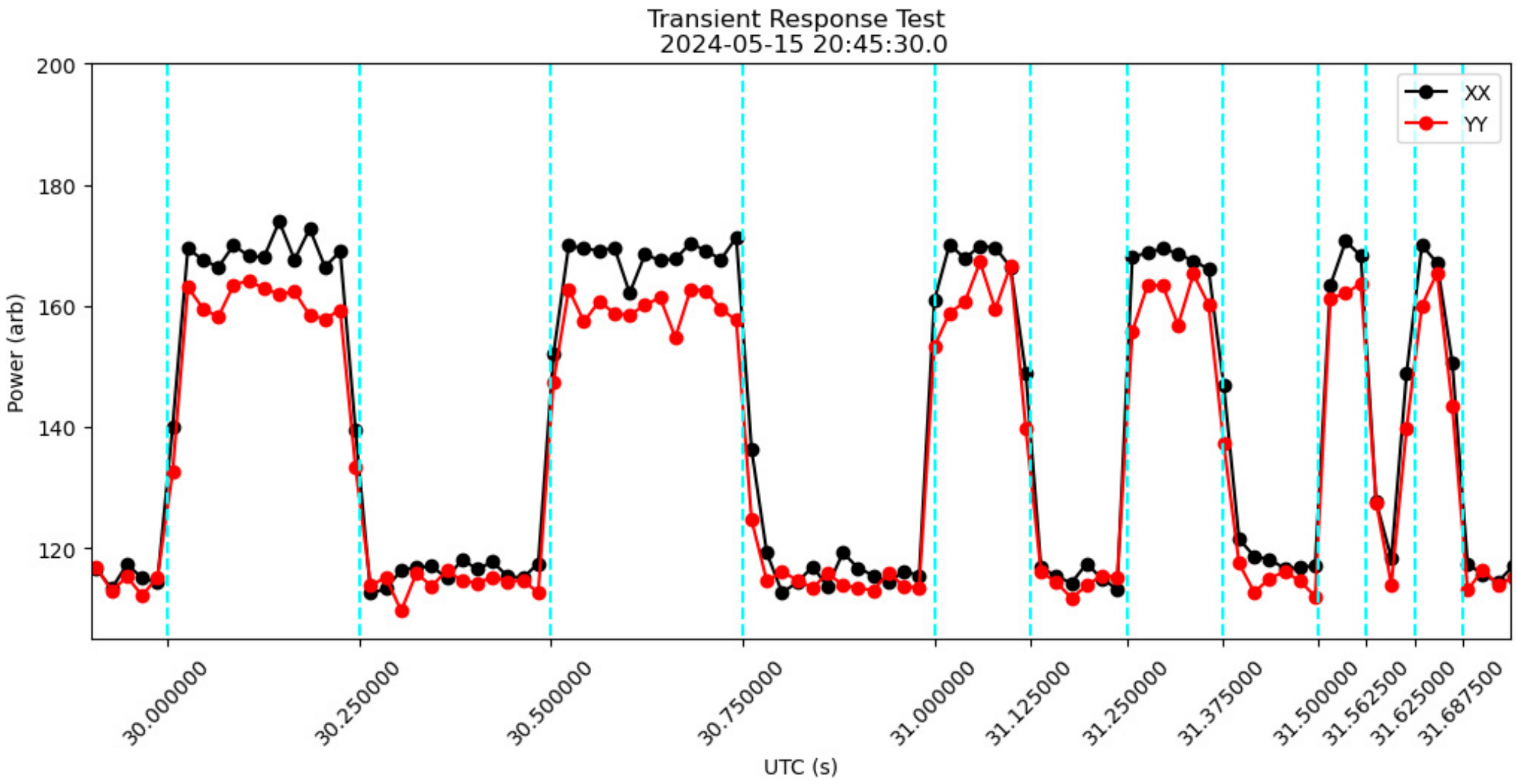

Fig. 7. (Color online) Station transient response on beam swaps between Cassiopeia A and the North Celestial Pole. Shown are on-off-on-off pointings between each source at geometrically decreasing intervals as described in Sec. 4.2. Vertical cyan dashed rules indicate MCS command timings of each source change for the intervals $\Delta t = 250$, 125, and 62.5 ms.

### 4.3. *Science observing validation*

A series of practical studies were conducted to illustrate the LWA-NA performance on observations typical for LWA single- and multi-station science. First, a comparison of pulsar B0950+08 observations taken at LWA-NA and LWA-SV, next, an ionospheric case study to triangulate Sporadic Es with three LWA stations located in New Mexico, and lastly, preliminary benchmarks for future LWA interferometry campaigns.

#### 4.3.1. *Pulsar observations*

Pulsar studies have been a key priority in LWA science with the LWA Pulsar Data Archive and over a dozen publications featuring LWA pulsar observations (Stovall *et al.*, 2015; Kumar *et al.*, 2022). This science has been led by the three LWA stations operating as single telescopes, but additional swarm stations, for example, could open new opportunities for pulsar observing below 100 MHz using the LWA. Here, we simply aim to show that this lighter-weight station architecture can detect pulsars at high signal-to-noise, and can be a valuable addition to existing or future LWA pulsar programs.

We used DRX beamforming observations at two LWA stations for this demonstration to compare the performance between standard- and swarm-sized LWA stations. PSR B0950+08 was observed using LWA-NA on August 29, 2024, and using LWA-SV on September 12, 2024, both for 20 min at 42.0 and 74.0 MHz tunings. Each tuning of the raw DRX data was reduced individually using the PRESTO pulsar toolkit to produce `prepfold` plots summarizing each observation (Ransom, 2001). Figures 8 and 9 show these plots for LWA-NA and Figs. 10 and 11 for LWA-SV.

Using the reported peak SNR of the pulsar, and the 74 MHz flux density measurement of $2.92 \pm 0.31\,Jy$ from the VLA Low Frequency Sky Survey (Cohen *et al.*, 2007), we can estimate the rms noise of the observation and SEFD for each station using the pulsar radiometer equation from Lorimer & Kramer (2005), simplified in Eq. (5). Where $\Delta\nu$ is the observed bandwidth, $t_{\rm acc}$ is the observation accumulation time, $W_{\rm eq}$ is the effective pulse width, and $P$ is the pulsar period. Below, we have substituted $\mathrm{SEFD} = G/T_{\rm sys}$ and $\Delta S = \mathrm{SNR}/S_\nu$, and assumed an ideal instrumental efficiency ($\beta = 1$) to resemble Eq. (4).

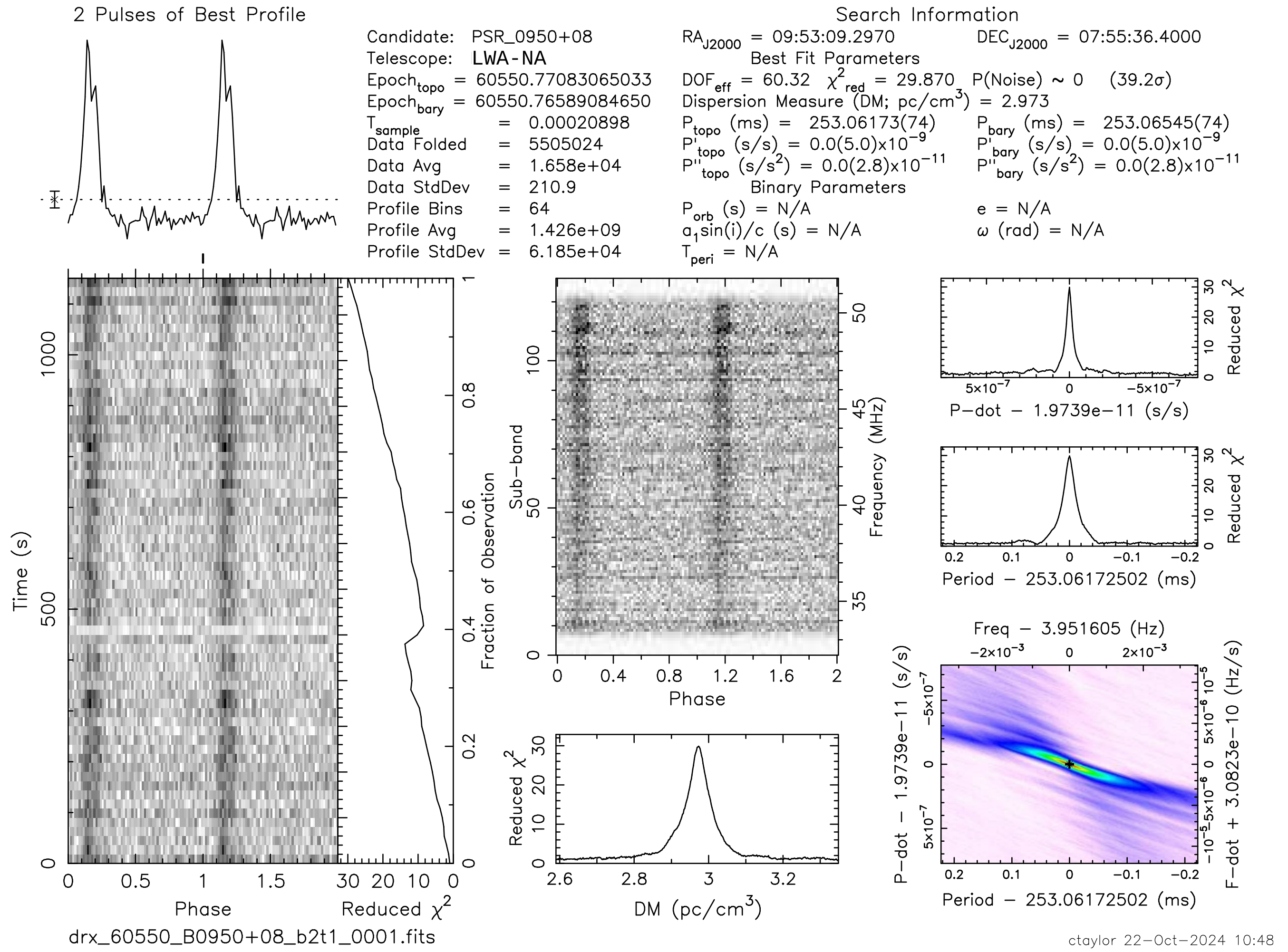

Fig. 8. (Color online) LWA – North Arm detection of pulsar B0950+08 during a commissioning observation on August 28, 2024, at 42 MHz. Pulsar data reduction and this `prepfold` plot were created using the PRESTO software package (Ransom, 2001).

$$\Delta S = \frac{\text{SEFD}}{\sqrt{n_p \Delta\nu t_{\text{acc}}}} \sqrt{\frac{W_{\text{eq}}}{P - W_{\text{eq}}}}. \tag{5}$$

Solving for SEFD and inputting the values from our observation with effective bandwidth $\Delta\nu_{\text{eff}} = 16.0\,\text{MHz}$, $t_{\text{acc}} = 1200\,\text{s}$ with a 10% duty cycle, and the number of polarizations $n_p = 2$, we can estimate the SEFD of each instrument. At LWA-NA, the peak SNR reported by PRESTO are $39.2\sigma$ at 42 MHz and $45.4\sigma$ at 74 MHz, resulting in an SEFD estimate of 43.7 kJy and 38.3 kJy, respectively. For LWA-SV, reported SNR values of $148.4\sigma$ and $145.7\sigma$ at 42 and 74 MHz correspond to SEFDs of 13.5 kJy and 16.7 kJy, respectively. Considering the differing RFI environments and instrumental effects at each site during the observations, we find these empirically derived SEFD measurements from B0950+08 to be consistent with those estimated in Sec. 3.4.

An upper limit to the SEFD is also regularly calculated at each LWA station using 10–15 min automated right ascension and declination cuts across Cygnus A during transit at 74 MHz as a reference. Since these station-monitored observations do not make corrections to remove the Galactic foreground, the value is best taken as an upper limit with an estimated error as high as 50% (Kumar *et al.*, 2025). Despite these error estimates, the SEFDs calculated during these station monitors at the time of this writing are 43.6 kJy at LWA-NA and 15.8 kJy at LWA-SV.

### 4.3.2. *Sporadic E*

One of the future goals of this experimental station is to establish tracking, or 'weather mapping', of Sporadic E (Es) structures in the local ionosphere.

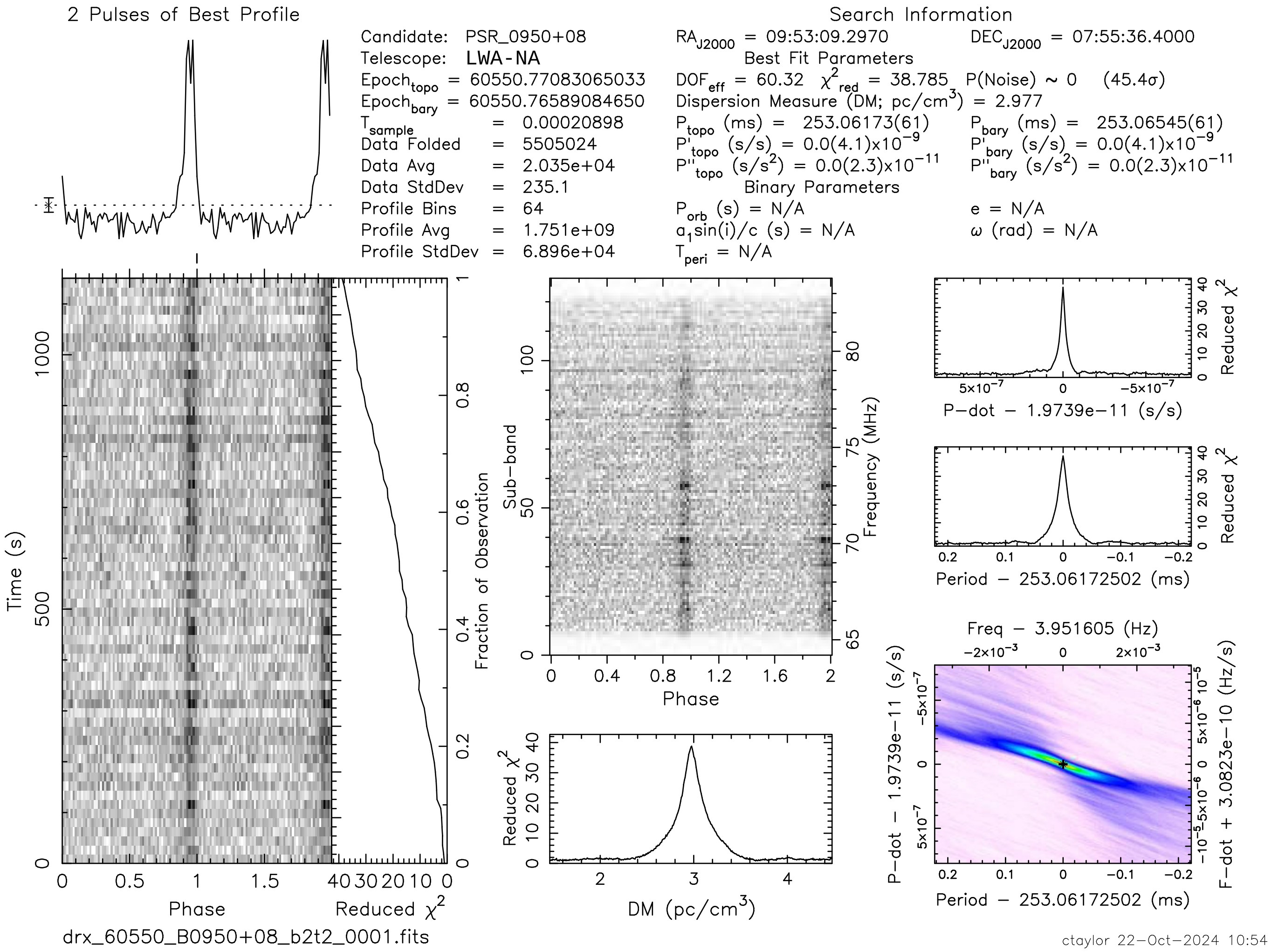

Fig. 9. (Color online) LWA – North Arm 74 MHz `prepfold` plot from the same observation seen in Fig. 8.

These structures are thin, transient enhancements to the electron density in the E-region of the ionosphere around 100 km altitude. Es regularly cause disturbances to RF communications and can reflect broadband ambient HF and VHF radio emissions over the horizon into low-frequency receivers. Obenberger *et al.* (2020) observed Es with two LWA stations and constrained the location of these ionosphere structures, and Obenberger *et al.* (2025) further mapped Es with confidence using LWA images and a digital ionosonde. Here we demonstrate that LWA-NA has the sensitivity and angular resolution to aid LWA1 and LWA-SV in studying these dynamic and transient structures (Obenberger *et al.*, 2020).

On August 25, 2024, a candidate event was identified using LWA-TV archived movies of a Es structure evolving over a ∼2 h period starting at 1600 UTC. Archival data were downloaded from each station, consisting of five-second all-sky snapshot maps, and then the background was subtracted using the LST model treatment of Obenberger *et al.* (2020) for data from LWA1 and LWA-SV. For LWA-NA a toy background model was created by taking a median filter over LST-matched data from August 24–26 on a pixel-by-pixel basis. Next, median images are subtracted from the LWA-NA archival data, and then background scrubbed images are averaged up into 5 min UTC bins. Following the procedures from Obenberger *et al.* (2025), accumulated images are masked to leave only connected structures brighter than five times the full image standard deviation, lowered from 10$\sigma$ to account for the slightly decreased sensitivity at LWA-NA compared to the 256-element stations, to be compared for Es localization. Previous models made for LWA1 and LWA-SV incorporate RFI flagging, image rejection, and station electronics temperatures to improve model stability, however, not enough stable archival data from LWA-NA exists currently to create a reliable model. These two background subtraction methods produce images of labeled connected structures, at 5 min intervals, that can be searched for Es using anti-coincidence at all three stations. In frequency, these images only cover a single 100 kHz channel in bandwidth at LWA1 (centered on the tuning center frequency), but LWA-SV and LWA-NA archival data are averaged into 3.3 and 3.25 MHz images, respectively, covering the full tuning bandwidth.

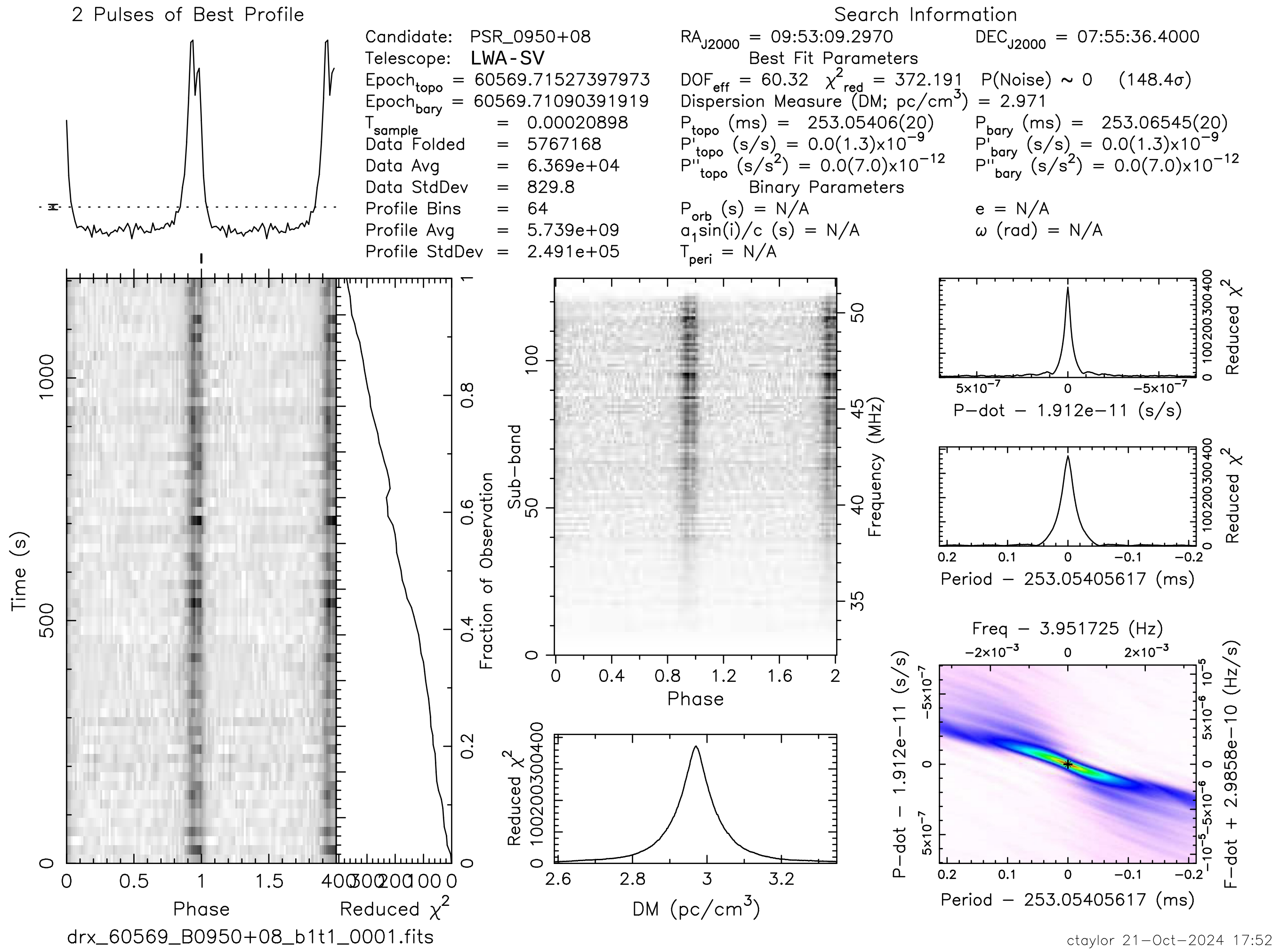

Fig. 10. (Color online) LWA – Sevilleta performance comparison observation of pulsar B0950+08 taken during commissioning on September 12, 2024, at 42 MHz. Pulsar data reduction and this `prepfold` plot were created using the PRESTO software package (Ransom, 2001). Note that the Sevilleta station contains 256-dipoles, yielding a tighter beam and better sensitivity than LWA – North Arm.

Detections of this Es structure are shown in panels (a), (b), and (c) of Fig. 12 at the three individual stations, with the propagating Es structure labeled due West in the images. Background subtracted images are projected onto a geographical map using tomography at the pixel level for each station, then assimilated into a look-up table to map pixels in LWA images onto a corresponding Latitude and Longitude grid. This procedure assumes that the Es are localized to an idealized altitude of 100 km and is described in more detail in Obenberger *et al.* (2025). The result, seen in panel (d) of Fig. 12, shows the combined intensity distribution when all three stations detect a Es in the same geographic grid cell. Due to the alignment of the source with the New Mexico LWA stations and the proximity of LWA-NA to LWA1 (∼ 20 km separation), the Es becomes elongated along the line of sight putting its approximate location over central Arizona. We track this source for approximately 1 h until it fades from the more narrowband images in LWA1. A better distribution of digital ionosondes and LWA Swarm sites would provide resolution, improved tracking, and an avenue to model our own local ionosphere for correction in beamformed observing.

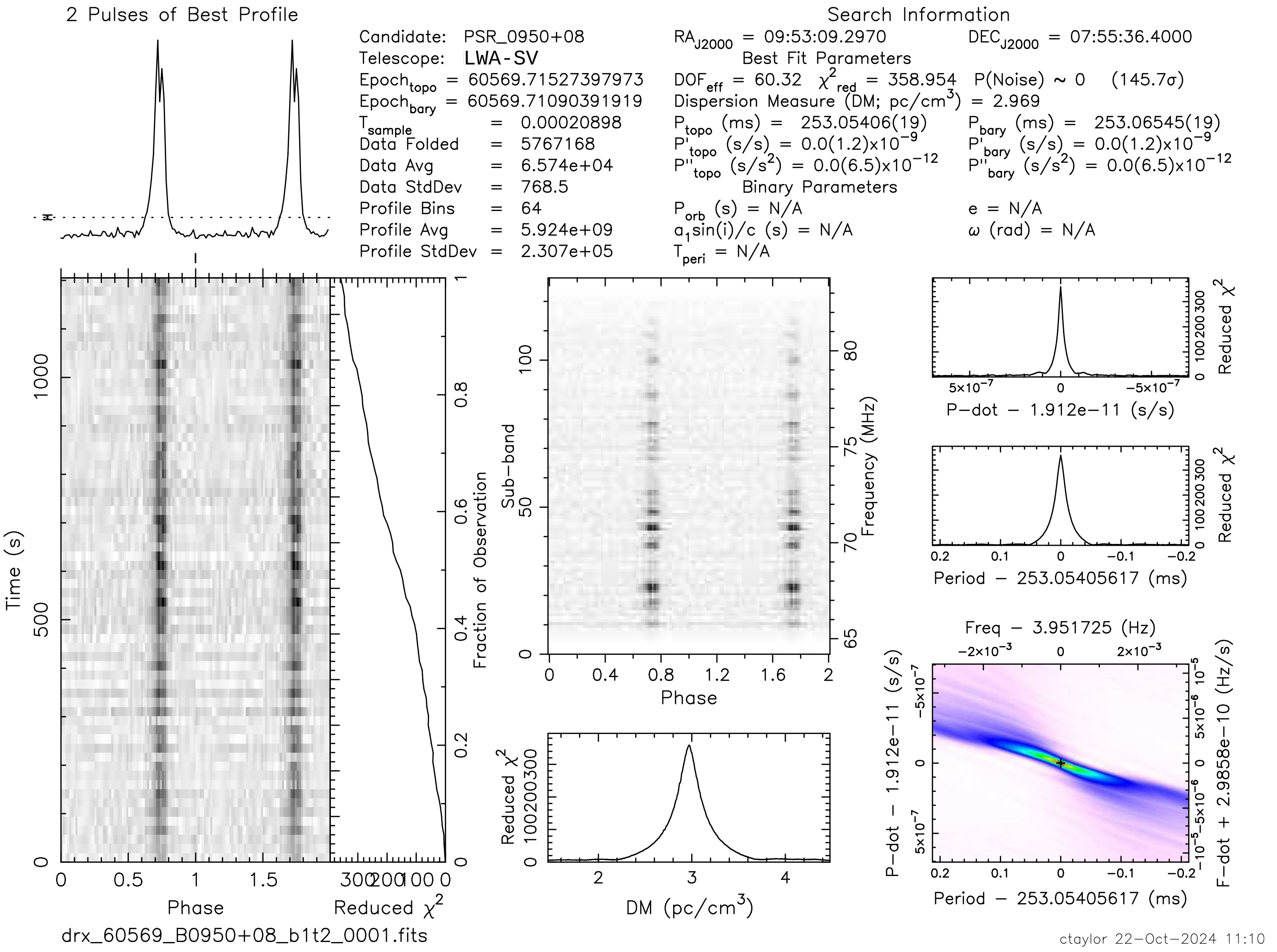

Fig. 11. (Color online) LWA – Sevilleta 74 MHz `prepfold` plot from the same observation seen in Fig. 10.

### 4.3.3. *LWA interferometry*

Davis *et al.* (2020) investigated the use of two LWA stations as a single baseline interferometer in observations of six nearby flaring UV Ceti type variable M-dwarfs. The two-element LWA finds a single faint flare from EQ Pegasi in this study, but identified the LWA Swarm as a progression to alleviate the calibration difficulties and to provide imaging capability. Here we present a set of plots from a few preliminary observations to illustrate that LWA Swarm stations are viable for long baseline interferometry. The $(u, v)$-coverage map of the three-element New Mexico LWA interferometer can be seen in Fig. 13 for test observations of the radio galaxy 3C295, a commonly used phase calibrator with a peak 74 MHz flux density of $\sim 120$ Jy. Reduction of this observation was done using NRAO Astronomical Image Processing System (AIPS; Greisen, 2003) and difmap (Shepherd, 1997). Provided also are cross-power spectra from a test observation of 3C123 on the unique baselines including LWA1 for each tuning in Fig. 14, where we can see stable fringes over two minutes in the phase plots above each baseline amplitude. For this heterogeneous array comprising two full stations and one mini station, the expected thermal noise estimate in the image plane at 74 MHz, assuming the ideal system temperature and effective area from Eqs. (1) and (2), is $\sim 5$ mJy rms noise in 8 h for 32 MHz of bandwidth (Ellingson *et al.*, 2009;

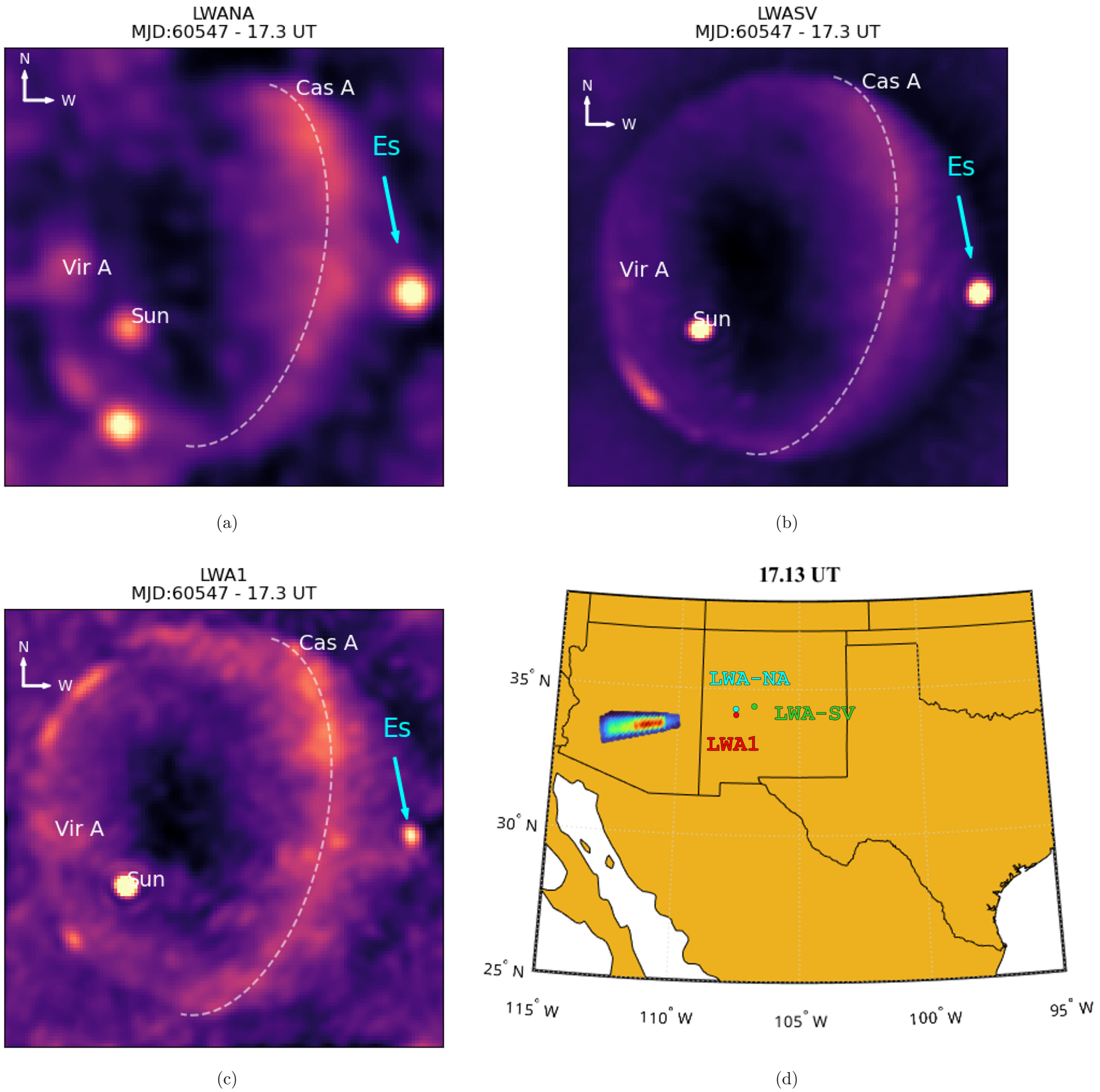

Fig. 12. (Color online) Demonstration of tomographic triangulation of Sporadic E (Es) regions in the ionosphere using the three LWA stations located in New Mexico. Sub-figures (a), (b) and (c) show a Es detection at each station located due West and labeled with an arrow in each plot. Panel (d) projects this structure using tomography onto a top-down map showing the structure is located over Arizona. The colormapping present in panel (d) indicates the summed intensity of each station in pixels where all telescopes can see the Es on the same projected geodetic Latitude and Longitude coordinate gridpoint. The location of stations is approximately to scale to aid in distinguishing LWA1 from LWA-NA since they are separated by only $\sim 20$ km.

Walker, 1989). If we can achieve this level of sensitivity after calibration, the LWA Swarm would be equipped deliver high SNR observating at low frequencies with more angular resolution than the upcoming SKA-LOW instument, while observing to substantially lower declinations than the LOFAR Array (Labate *et al.*, 2017; de Gasperin *et al.*, 2021).

## 5. Discussion

LWA-NA represents the first step in expanding the LWA Collaboration with an LWA Swarm, a more powerful low-frequency, long-baseline interferometer at mid-latitudes. Its implementation provides a smaller footprint to test experimental changes or future upgrades to the LWA operations architecture. LWA-NA eschews several custom-designed components of previous LWA stations in New Mexico, namely the electronics shelter and original digital architecture described herein, to improve station cost efficiency, availability of parts, and flexibility for future modifications. For example, LWA1 and LWA-SV house excess cable in a vault embedded in the ground adjacent to the SEP on each station. To mitigate this cost and improve protection from wildlife (primarily rodents) at LWA-NA, we cut an approximately $10 \times 10 \times 2$ ft. auxiliary trench, where the excess cable was spooled up before being buried following completion of construction. Similarly, wildlife intrusion at antennas was found to be a common problem at LWA1, LWA-SV, and OVRO-LWA. To prevent wildlife intrusions at LWA-NA that could damage cables or electronics, extra care was taken when sealing cable entry ports on antenna masts using screws and PVC cement.

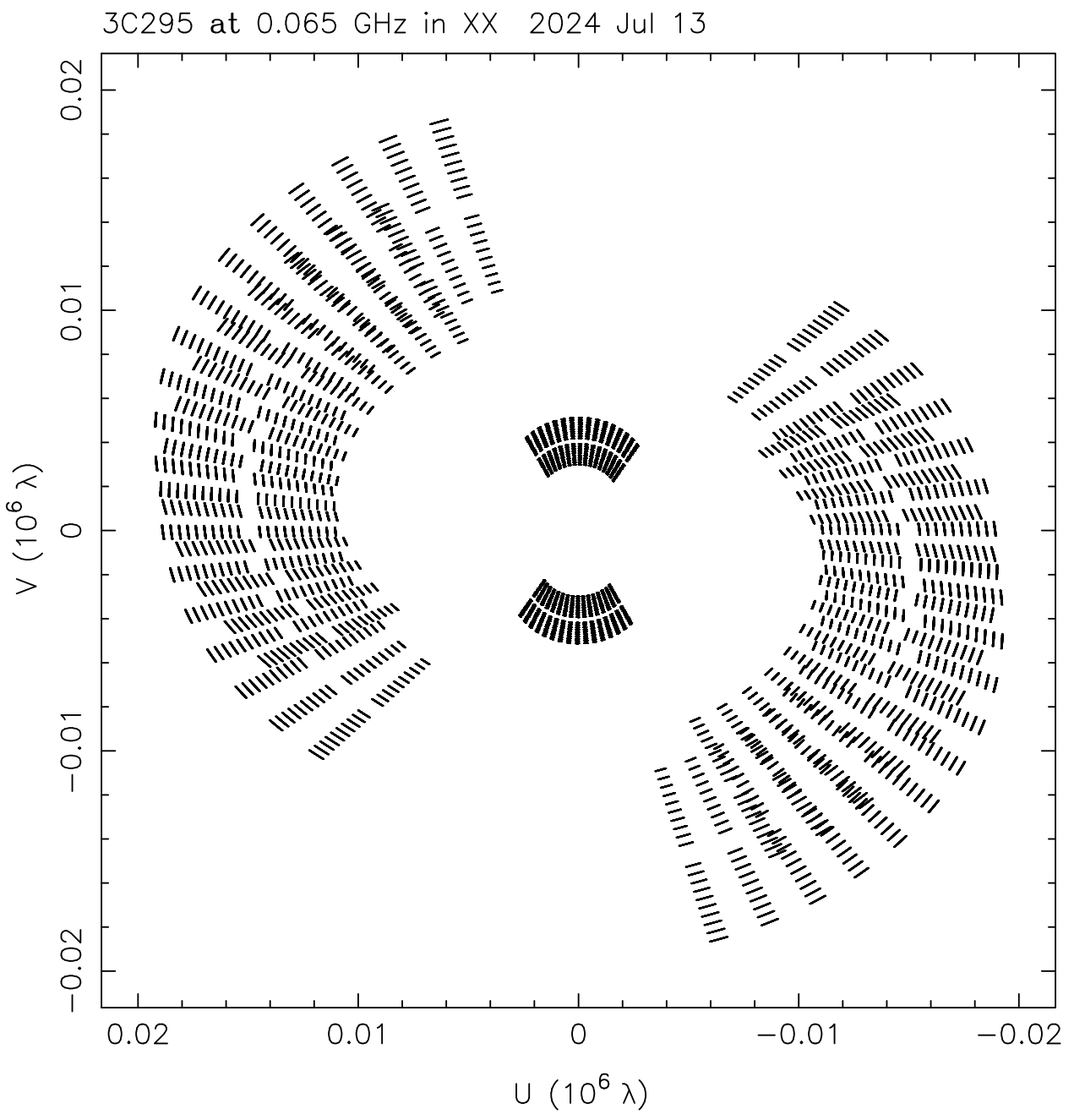

Fig. 13. $(u, v)$-coverage map from an 8 h preliminary interferometry observation of 3C295 at 55 MHz (pictured) and 74 MHz tunings using the three New Mexico LWA stations (LWA1, LWA-SV, and LWA-NA) on July 13, 2024. The image was produced using Difmap after data reduction in AIPS (Shepherd, 1997; Greisen, 1988).

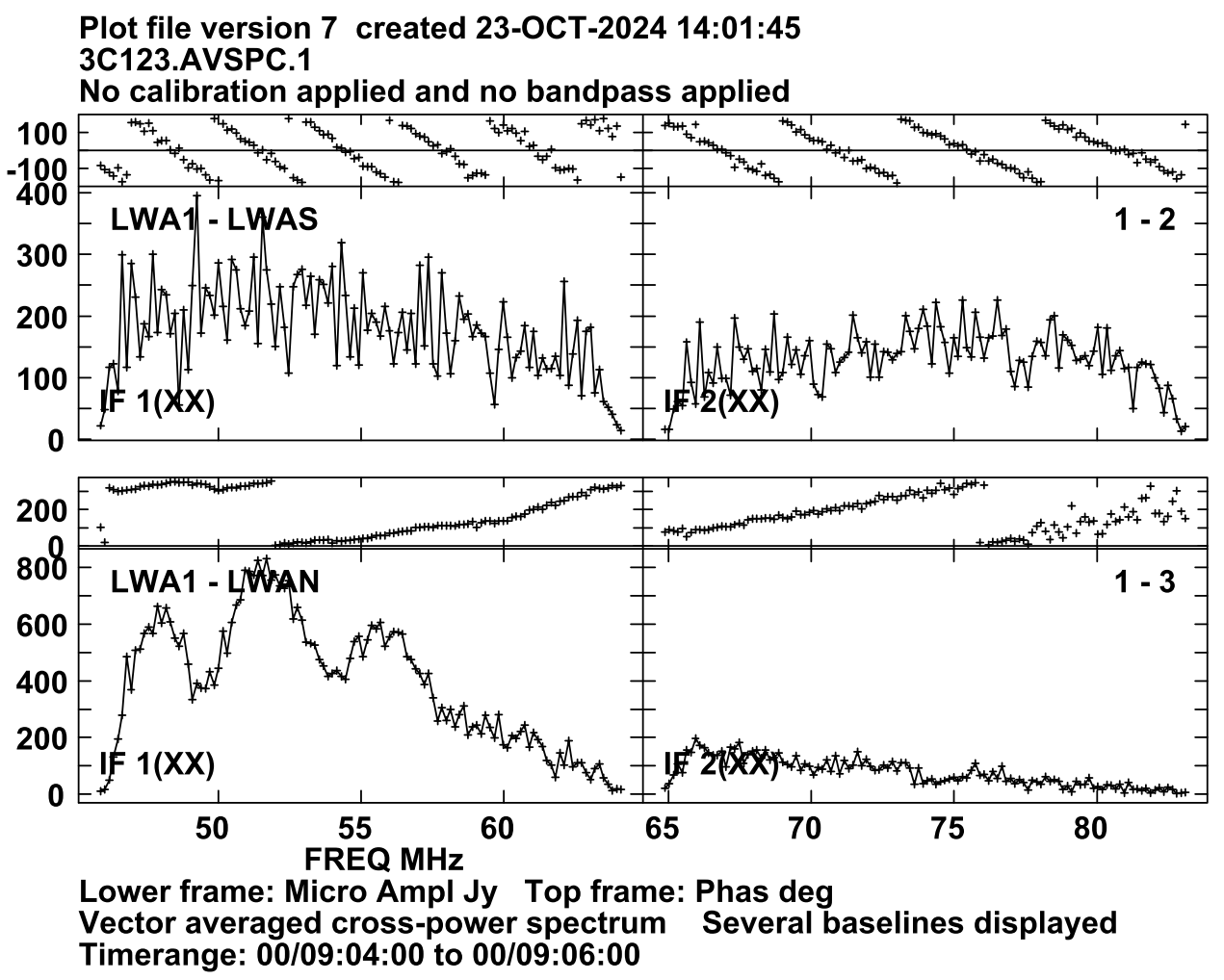

Fig. 14. 2 min cross-power spectrum for LWA1 (1) referenced baselines to LWA-SV (2) and LWA-NA (3) during a test interferometry observation of 3C123 on October 3, 2024.

The design and commissioning of this swarm station process has been documented in the LWA Memo series to aid in streamlining future installations, particularly for two additional 64-element swarm stations already funded in collaboration with the Arizona State University and Texas Tech University. On-site developments are already underway for these two new swarm stations to be located at Meteor Crater National Landmark in Northern Arizona (LWA-MC) and at the Comanche Springs Astronomy Campus near Crowell, TX (LWA-CS; Dowell *et al.*, 2023). In the development of LWA-NA, we spent considerable effort to validate individual signal paths with degraded performance caused by damaged or failed FEE components, connector damage from cable strain or incorrect installation, induced damage to the ARX channels corresponding to a given antenna, or digital channel anomalies. We made the applicable repairs to these antennas, but 7% of channels still contain transient spectral errors caused by an unknown issue on individual SNAP2 channels. Due to the cost and lead time to acquire replacement SNAP2 units, we are developing a more accessible digitizer format for two future stations using a Xilinx HW-Z1-ZCU102 unit. To address the challenges encountered in the development of LWA-NA, these two future swarm stations will incorporate a redesigned backend including the next revision ARX boards (based on failure analysis from OVRO-LWA and LWA-NA), a new digitizer configuration to accommodate the Xilinx FPGAs, and use cut-to-length coaxial cable with a cable strain relief system inside the SEP.

Finally, a post-commissioning focus will be surveying a selection of bright active galactic nuclei to provide an early set of calibrator sources for LWA interferometry. Candidate calibrators were selected from a combination of the VLA Calibrator List and VLA Low-Frequency Sky Survey Redux catalog (Lane *et al.*, 2014), with observations currently underway. An analysis of these sources as calibrators for the LWA Swarm will be the topic of a sequel paper to this publication. As new stations join the LWA Swarm over the course of 2025–2026, we expect to see immediate improvements to multi-station science, especially in observing the dynamic ionosphere and interferometry applications.

## Acknowledgments

We acknowledge the efforts of the following students who helped in the construction of the LWA-NA station: UNM students Lily Wood, Sean Kinney, Dylan Henderson, and Juan Ramirez, Hillsdale College students Evan Anthopoulos, Riley Hamilton, Paige Lettow, Joseph Petullo, Liam Swick, and Nathan Sibert, and Rochester Institute of Technology students Olivia R. Young and Jackson Hebel. T.E.D. was supported by Hillsdale College sabbatical and summer leave funding, by NSF Astronomy and Astrophysics Grant (AAG) award number 2009468, and, along with the NANOGrav Student Teams of Astrophysics ResearcherS (STARS) program, was supported by National Science Foundation (NSF) Physics Frontiers Center award No. 2020265. T.E.D. acknowledges the Hillsdale College Division of Natural Sciences and the Hillsdale College Dept. of Physics for additional travel funding.

We also thank NRAO and the VLA staff for providing on-site assistance in the construction of the LWA-NA swarm station. We are grateful to the University of Texas—Rio Grande Valley for the repurposing of Low-Frequency All-Sky Monitor equipment from the North Arm site. Construction of the LWA has been supported by the Office of Naval Research under Contract N00014-07-C-0147. Support for operations and continuing development of the LWA is provided by the Air Force Research Laboratory and the National Science Foundation under grants AST-2107845. This research was sponsored in part by the Air Force Office of Scientific Research (AFOSR) Lab Task 23RVCOR002. Support for construction of the LWA-NA station was provided by the Air Force Office of Space Research Lab Task 23RVCOR002.

The views expressed are those of the authors and do not reflect the official guidance or position of the United States Government, the Department of Defense or of the United States Air Force. The appearance of external hyperlinks does not constitute endorsement by the United States Department of Defense (DoD) of the linked websites, or the information, products, or services contained therein. The DoD does not exercise any editorial, security, or other control over the information you may find at these locations.

## ORCID

C. A. Taylor https://orcid.org/0009-0006-5879-3105
J. Dowell https://orcid.org/0000-0003-1407-0141
G. B. Taylor https://orcid.org/0000-0001-6495-7731
K. S. Obenberger https://orcid.org/0000-0002-6702-4264
S. I. Chastain https://orcid.org/0000-0003-3507-335X
L. E. Cordonnier https://orcid.org/0009-0009-8806-479X
P. Kumar https://orcid.org/0000-0003-0101-1986
S. Bruzewski https://orcid.org/0000-0001-7887-1912
T. E. Dolch https://orcid.org/0000-0001-8885-6388